\documentclass[journal]{IEEEtran}
\usepackage[T1]{fontenc}
\usepackage[utf8]{inputenc}

\usepackage{orcidlink}
\usepackage{cite}
\usepackage{subcaption}
\usepackage{tabularx} 
\usepackage{amsmath}
\usepackage{graphicx}
\usepackage{url}
\usepackage{xcolor}

\begin{document}
%
\title{First Experimental Results of Space Object Detection Using the InteRFeX Array}
%
%
%


\author{Bhaskar~Ahuja\,\orcidlink{0009-0001-8126-2031},
        Thomas~Crowther,
        Steven~Tingay\,\orcidlink{0000-0002-8195-7562},
         Andrew~Guidi, Luca~Gentile
        and~Marco~Martorella\,\orcidlink{0000-0002-8985-5069},~\IEEEmembership{Fellow,~IEEE}

\thanks{Bhaskar Ahuja is with the Department of Physics, University of Trento, Trento, Italy, and also with the Radar and Surveillance Systems (RaSS) National Laboratory of CNIT, 56124 Pisa, Italy (e-mail: bhaskar.ahuja@unitn.it).}
\thanks{Thomas Crowther and Steven Tingay are with the International Centre for Radio Astronomy Research, Curtin University, Bentley, WA, Australia (e-mail: thomas.crowther@postgrad.curtin.edu.au; s.tingay@curtin.edu.au).}%
\thanks{Andrew Guidi is with Nova Systems, Level 3, 169 Pirie Street, Adelaide, SA 5000, Australia (e-mail: Andrew.guidi@novasystems.com).}
\thanks{Luca Gentile is with the Radar and Surveillance Systems (RaSS) National Laboratory of CNIT, 56124 Pisa, Italy (e-mail: luca.gentile@cnit.it).}%
\thanks{Marco Martorella is with the Microwave Integrated Systems Laboratory (MISL), University of Birmingham, Birmingham, U.K., and also with the Radar and Surveillance Systems (RaSS) National Laboratory of CNIT, 56124 Pisa, Italy (e-mail: m.martorella@bham.ac.uk).}
}

%
%

\markboth{IEEE TRANSACTIONS ON RADAR SYSTEMS \ }%
{Shell \MakeLowercase{\textit{et al.}}: Bare Demo of IEEEtran.cls for IEEE Journals}
%



\maketitle

\begin{abstract}

This paper presents the first experimental validation of InteRFeX, a non-uniform  receive only (passive) array located at the Nova Systems Space Precinct in Peterborough, Australia, designed for space domain awareness applications. It operates over a frequency range of $70~\text{to}~300~\mathrm{MHz}$ and is capable of functioning in a range of observational modes, including as a coherent passive bistatic radar for space surveillance, which is the focus of this work. InteRFeX exploits commercial broadcast transmissions as illuminators of opportunity, combining cross-ambiguity function processing for bistatic range and range-rate retrieval with MUSIC-based direction-of-arrival (DoA) estimation for angular localization. The International Space Station (ISS) serves as the experimental target, with measurements evaluated against Two-Line Element (TLE) orbit predictions. {{Taking advantage of the dual-polarized array, all four reference--surveillance polarization combinations are evaluated, revealing differences in detection rate and mean signal-to-noise ratio (SNR) over the observation pass, with bistatic-range RMSEs of approximately $1~\mathrm{km}$ relative to the TLE-derived reference.} Independently, the MUSIC-based DoA estimates exhibit a mean angular separation of approximately $0.45^\circ$ from the TLE-derived reference directions. {The angular deviations are further decomposed into along-track and cross-track components, enabling a more detailed characterization of the DoA offsets relative to the TLE-derived reference trajectory.} The processing framework also includes a first-order bistatic link-budget analysis to estimate the expected SNR over the observation geometry. These results demonstrate the feasibility of passive-radar space domain awareness using radio-interferometric array technology and establish the proposed signal-processing pipeline as a basis for a dedicated coherent passive-radar mode within InteRFeX.}

\end{abstract}


%
\IEEEpeerreviewmaketitle

\section{Introduction}
\IEEEPARstart{T}{he} rapid growth of operational satellites and debris in Low Earth Orbit (LEO) has increased the demand for reliable space domain awareness (SDA) capabilities to prevent collisions and support safe space operations \cite{klinkrad2004esa}. Resident space object surveillance has traditionally relied on active radar and optical systems. Among these, active radars provide all-weather coverage but require high transmit power and substantial infrastructure, which limits deployment scalability and detection performance for small objects at long ranges \cite{mehrholz1997radar}. While optical  systems operate passively, they depend strongly on illumination geometry and atmospheric conditions, reducing their effectiveness for continuous monitoring \cite{hobson2016catalogue}. These limitations motivate the exploration of alternative sensing approaches capable of persistent and cost-effective observation of LEO objects.

Passive radar has emerged as a promising candidate for SDA because it exploits existing transmitters of opportunity and does not require a dedicated high-power transmitter \cite{palmer2012dvb}. Among passive sensing approaches, radio astronomy facilities are particularly attractive as receivers due to their extremely high sensitivity, originally designed for detecting weak cosmic signals \cite{ahuja2025role}. Early studies demonstrated bistatic space object detection by pairing large single-dish radio telescopes with powerful transmitters, such as the combination of the Effelsberg telescope and the Tracking and Imaging Radar (TIRA) in Germany \cite{letsch2009first}. While effective, these configurations are constrained by narrow beamwidth and limited sky coverage, making them unsuitable for wide-area surveillance.

To overcome these limitations, attention has shifted toward array-based radio telescopes, which combine high sensitivity with electronic beam steering and large instantaneous field of view \cite{ahuja2025role}. Such arrays provide scalable receiver architectures, operate over wide bandwidths, and can simultaneously support multiple observation directions. These characteristics make them well suited for passive radar operation using non-cooperative transmitters such as FM radio and digital television broadcasts. { Several large interferometric arrays, particularly the Low Frequency Array (LOFAR \cite{van2013lofar}) in Europe and the Murchison Widefield Array (MWA \cite{tingay2013murchison}) in Australia  have demonstrated feasibility for SDA applications. Over the past several years, LOFAR has been extensively investigated as a passive radar receiver for space situational awareness using terrestrial digital broadcast transmitters as illuminators of opportunity (IoO) \cite{jkedrzejewski2025emerging,jkedrzejewski2022first}. Initial studies demonstrated the feasibility of passive space object detection using digital audio broadcasting (DAB) transmissions \cite{malanowski2021passive}. Subsequent experimental campaigns transitioned to higher-bandwidth digital video broadcasting terrestrial (DVB-T) illuminators, enabling improved bistatic range resolution and the successful detection of the International Space Station (ISS) \cite{jkedrzejewski2022experimental}. In contrast, MWA has successfully detected numerous LEO objects using FM radio transmissions as IoO \cite{tingay2013detection,prabu2020development,prabu2020low,prabu2022improved,prabu2023demonstration}. Despite their relatively limited bandwidth, the widespread availability and high transmitted power of FM broadcast signals make them attractive illuminators for passive radar. Building on these advantages, multistatic observation campaigns with the MWA demonstrated that fusing measurements from multiple FM transmitters significantly improved orbit determination accuracy by reducing position and velocity uncertainties \cite{hennessy2023surveillance,hennessy2022establishing}.} Similar concepts are also being investigated with the Meer Karoo Array Telescope (MeerKAT) in South-Africa and the BIRALES system in Italy \cite{dhondea2019mission,losacco2019multibeam}. Beyond academic facilities, commercial passive radar systems exploiting broadcast transmitters have begun to emerge, further indicating the operational potential of this approach. For instance, Silentium Defence’s Maverick exploits FM broadcast transmissions to detect objects and support orbit estimation in LEO \cite{finch2024passive}.

Despite these significant advances, several practical challenges remain. While both the LOFAR and MWA studies have demonstrated the feasibility of passive space surveillance using low-frequency radio astronomy infrastructure, these facilities are primarily designed for astronomical observations, potentially limiting their availability for continuous passive radar operations. In addition, conventional passive radar architectures typically employ a dedicated reference antenna, requiring additional hardware and precise synchronization between the reference and surveillance channels. Recent studies using the MWA and LOFAR have demonstrated that both channels can instead be formed from a common antenna array \cite{palmer2017surveillance,sameantennalofar,jkedrzejewski2024experimental}. While the LOFAR implementation partitions the array into separate reference and surveillance sub-arrays, the MWA approach employs the full array for both channels, avoiding a fixed aperture division and supporting multiple reference beams.

The present work addresses these limitations through the development of InteRFeX, a dedicated array-based receive-only (passive) system for SDA. The system adapts radio astronomy array concepts for surveillance operation, enabling repeatable measurements without reliance on astronomical observing schedules. {Furthermore, inspired by the full-array beamforming approach demonstrated with the MWA \cite{palmer2017surveillance}, InteRFeX employs a unified digital beamforming architecture in which the reference and surveillance beams are formed simultaneously. This avoids a fixed aperture division and removes the need for a dedicated external reference antenna and inter-channel synchronization, while supporting multiple reference beams for future multistatic operation. {In addition, each array element supports dual-polarization reception, enabling simultaneous acquisition of two orthogonal polarization components of both the direct-path and target-scattered signals.} This capability enables flexible polarization combinations in post-processing and helps accommodate the mixed-polarization nature of modern broadcast transmissions, improving robustness of reference formation and target detection under practical operating conditions. Although recent studies have begun to consider polarization information in passive radar processing, its systematic exploitation in FM broadcast based SDA with large dual-polarized arrays remains at an early stage \cite{polarimetric_passive_radar_fabiola,filippini2020polarimetric,jkedrzejewski2025influence}. 

To provide an initial experimental validation of InteRFeX for passive LEO surveillance, a measurement campaign was conducted in November 2025 using the ISS as the target. The campaign evaluated the recovery of bistatic range, range rate, and angular information from the received target echoes. The main contributions of this paper are summarized as follows:

\begin{itemize}
    \item {Development of a signal-processing pipeline for passive radar operation of InteRFeX, enabling passive detection of space objects while forming both reference and surveillance channels with the same receive array through unified digital beamforming.}

     \item {First experimental demonstration of InteRFeX as a passive bistatic radar for detecting the ISS, achieving angular localization through direction-of-arrival estimation and extracting bistatic range and Doppler information using the cross-ambiguity function, with validation against its TLE-predicted trajectory.}
 
   \item A polarization-dependent performance analysis exploiting dual-polarization reception, evaluating the influence of the {reference–surveillance X/Y channel combinations} on detection rate, signal-to-noise ratio (SNR), and bistatic measurement performance under practical operating conditions.

\end{itemize}

The remainder of this paper is organized as follows. Section~\ref{bg} presents the system description, including the beamforming architecture and signal-processing pipeline. Section~\ref{results} describes the experimental setup, presents the results, and discusses the key findings. Sections~\ref{conclusions} and~\ref{future_work} present the conclusions and outline directions for future work.

\section{Background}\label{bg}
\subsection{System Description }\label{system}

InteRFeX is a low-frequency passive array located at the Nova Systems Space Precinct in Peterborough, South Australia. It is a dedicated SDA instrument with the intention of enabling continuous data capture and monitoring of the Space Domain. InteRFeX has an operational frequency range of $70\text{--}300~\mathrm{MHz}$, and can observe in multiple operating modes, including as a radio interferometer, and, as described in this paper, as a passive bistatic radar receiver. The array consists of 16 elements, called Tiles, situated in a randomly perturbed grid over a total area of approximately $35~\mathrm{m} \times 35~\mathrm{m}$. {Each Tile is an analogue beamformed sub-array, similar in design to an MWA tile \cite{tingay2013murchison}. It consists of $16$ dual-polarized dipole antennas arranged in a $4\times4$ grid with a $1.1~\mathrm{m}$ inter-element spacing.} The dipoles are located on a $50~\mathrm{mm} \times 50~\mathrm{mm}$ steel mesh base with a $5~\mathrm{m} \times 5~\mathrm{m}$ extent, which acts as a ground plane for the Tile. {Each dipole pair contains low-noise amplifiers (LNAs) and provides two
orthogonal analogue outputs, denoted X and Y, corresponding to the
East--West and North--South linear-polarization orientations, respectively.} Fig.~\ref{fig:array_photo} shows the InteRFeX array as deployed at the Nova Systems Space Precinct in Peterborough, Australia. The non-uniform spacing between the 16 Tiles is evident across the aperture, while the enlarged views highlight the structure of an individual dual-polarized crossed bow-tie antenna and the corresponding $4\times4$ Tile configuration.

\begin{figure}[!t]
    \centering
    \includegraphics[width=1\linewidth]{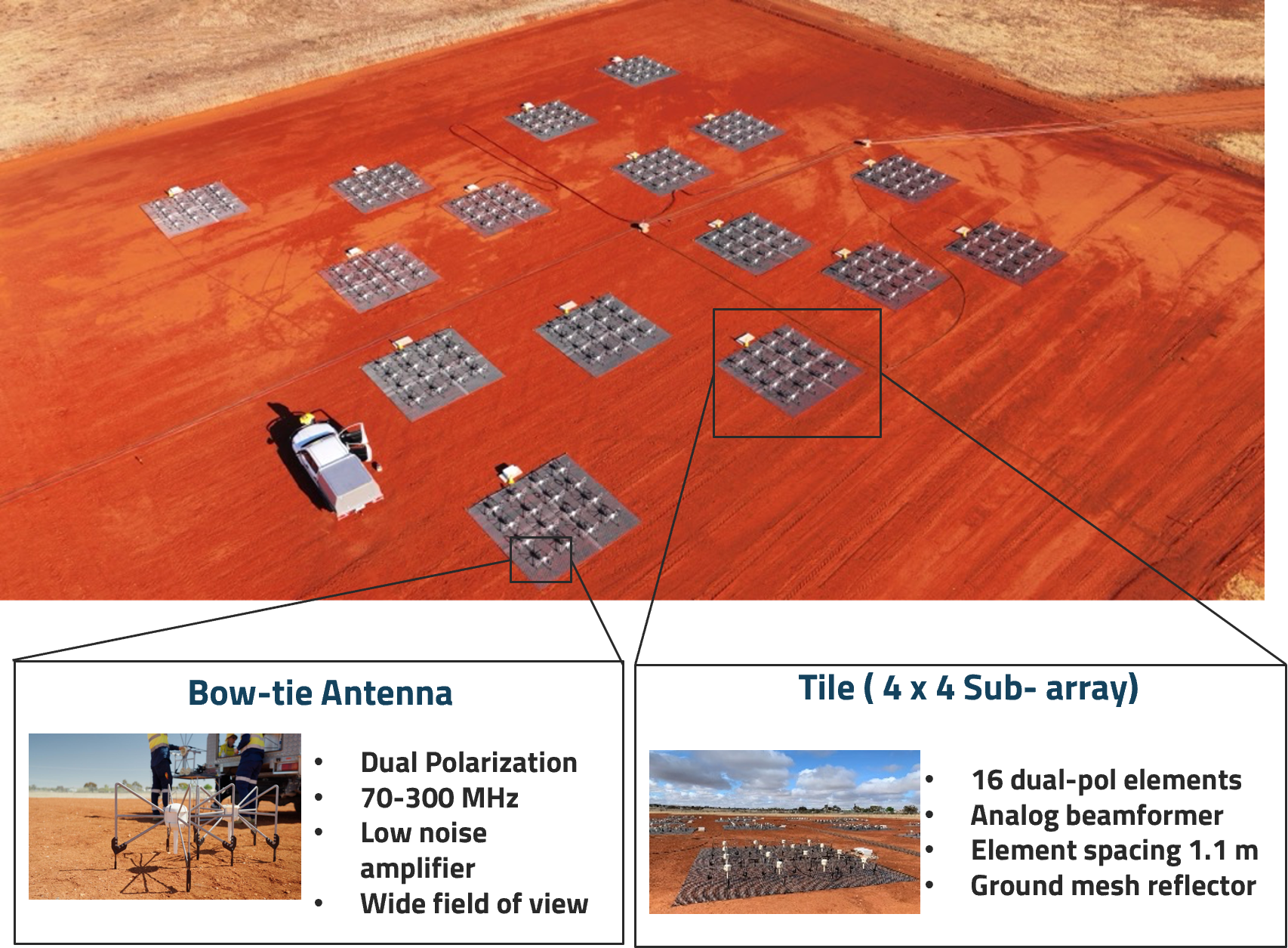}
   \caption{{Wide-field view of the InteRFeX antenna system at the Nova Systems Space Precinct, Peterborough, Australia, with enlarged views of a dual-polarized crossed bow-tie antenna element (left) and a Tile (right) comprising 16 elements.}}
    \label{fig:array_photo}
\end{figure}

{ The receive array employs a hierarchical analogue--digital beamforming architecture. At the Tile level, the signals from the $4\times4$ crossed bow-tie elements are combined using an analogue true-time-delay beamformer, yielding one X-polarized and one Y-polarized output per Tile. The simulated radiation patterns are represented in the direction-cosine, or $u-v$, domain, defined as
\begin{equation}
u=\cos(\theta)\sin(\phi), \qquad
v=\cos(\theta)\cos(\phi),
\label{eq:direction_cosines}
\end{equation}
where $\theta$ and $\phi$ denote the elevation and azimuth angles, respectively. At the operating frequency of $107.9~\mathrm{MHz}$, the simulated zenith-pointed Y-polarized Tile beam exhibits half-power beamwidths (HPBWs) of approximately $31.47^\circ$ and $30.23^\circ$ along the $u$- and $v$-axes, respectively, as shown in Fig.~\ref{fig:tile_pattern}. The near-similar beamwidths indicate an approximately symmetric broadside response and define the instantaneous primary field of view available to the subsequent digital processing stage. The analogue Tile beam can be steered in both azimuth and elevation, although its sensitivity varies with steering direction. Owing to the radiation pattern of the crossed-bowtie elements above the ground screen, the sensitivity is maximum near zenith and decreases toward lower elevation angles, with a more rapid roll-off below approximately $30^\circ$ elevation.

\begin{figure}[!t]
    \centering
    \includegraphics[width=1\linewidth]{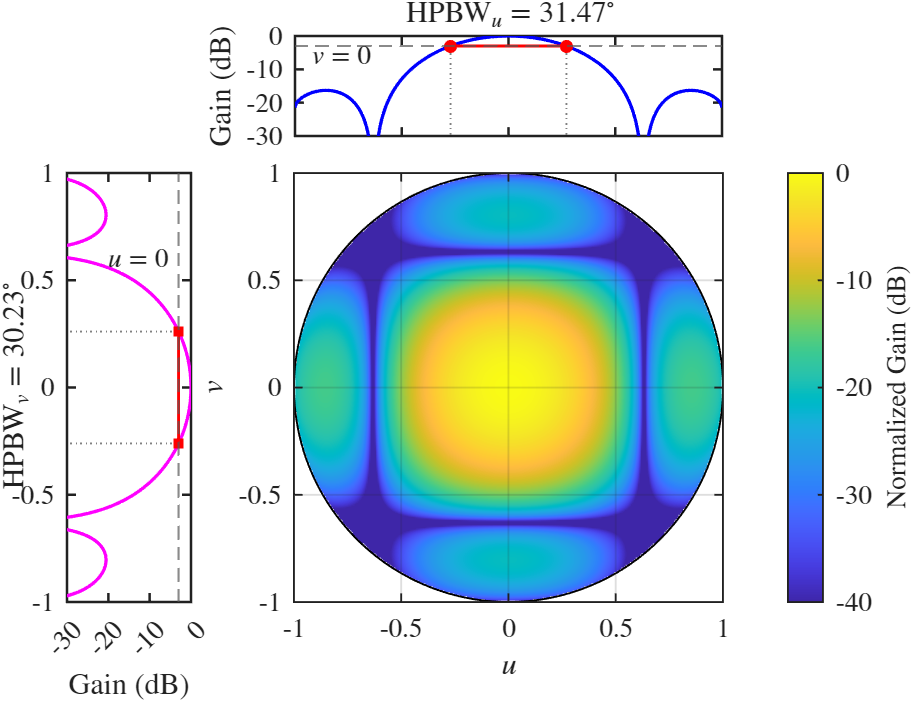}
\caption{{Normalized Y-polarized beam pattern of a single Tile at $107.9~\mathrm{MHz}$ for zenith pointing. The central panel shows the pattern in the visible $u$--$v$ region, while the top and left panels show the $v=0$ and $u=0$ cuts with HPBWs of $31.47^\circ$ and $30.23^\circ$, respectively.}}
    \label{fig:tile_pattern}
\end{figure}

\begin{figure}[!t]
    \centering
    \includegraphics[width=1\linewidth]{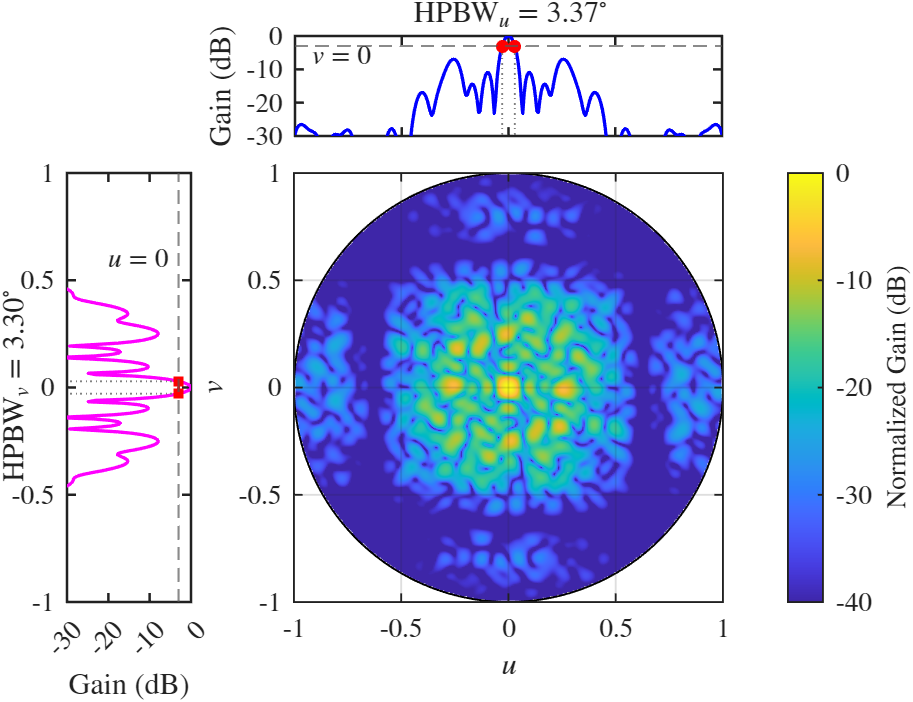}
\caption{{Normalized Y-polarized beam pattern of the full 16-tile array at $107.9~\mathrm{MHz}$ for zenith pointing. The central panel shows the pattern over the visible $u$--$v$ region, while the top and left panels show the $v=0$ and $u=0$ cuts, with HPBWs of $3.37^\circ$ and $3.30^\circ$, respectively.}}
    \label{fig:array_pattern}
\end{figure}

Consequently, the resulting X- and Y-polarized Tile outputs are digitized by individual Airspy software-defined radios (SDRs) at a rate of $5$ Msamp/s as $16$-bit real samples, providing a system bandwidth of $2.5~\mathrm{MHz}$. Subsequently, digital beamforming is performed on the digitized outputs from all 16 Tiles to synthesize independent receive beams towards the illuminator and the target. Since the analogue Tile beam remains fixed at zenith throughout the observation, all subsequent steering of the reference and surveillance beams toward the illuminator and target, respectively, is performed in the digital domain. Although a fully digital architecture at the individual-dipole level would provide greater beamforming flexibility, the analogue Tile primary beam provides a half-power field of view of approximately 740 degree square at $107.9~\mathrm{MHz}$, corresponding to an equivalent angular diameter of approximately 31$^\circ$. Many narrower digital beams can therefore be formed simultaneously within this Tile field of view for the detection and tracking of multiple targets, while maintaining practical system cost and complexity.

Fig.~\ref{fig:array_pattern} illustrates the resulting array pattern at the selected operating frequency of $107.9~\mathrm{MHz}$, where coherent combination of the 16 spatially distributed Tile outputs produces a substantially narrower broadside beam, with HPBWs of approximately $3.37^{\circ}$ and $3.30^{\circ}$ along the two principal angular directions.  These values correspond to the zenith-pointed configuration and vary with both the operating frequency and the beam-steering direction. As the beam is steered away from zenith, the effective projected aperture decreases, resulting in direction-dependent main-beam broadening, particularly in elevation toward the horizon.}

Another important aspect is the non-uniform distribution of the $16$ Tiles over a large aperture, resulting in sparse spatial sampling. This irregular layout suppresses periodic grating lobes by redistributing their energy into weaker, diffuse sidelobes, while the large aperture preserves a narrow main lobe~\cite{jiwani2013square}. Moreover, FM broadcast services operate in the VHF band from approximately $88$--$108~\mathrm{MHz}$, with high-power transmitters providing continuous illumination over large coverage areas \cite{ACMA_SpectrumPlan2025}. DAB services in Australia operate within VHF Band~III, approximately $174$--$230~\mathrm{MHz}$, and similarly provide continuous broadcast signals with well-defined waveform characteristics \cite{ACMA_SpectrumPlan2025}. The operating bandwidth of InteRFeX encompasses both FM and DAB/DAB+ broadcast bands, making these services suitable for use as IoO.

\subsection{Signal processing}\label{SP}

The operation of a passive bistatic radar depends on both the transmitter--target--receiver geometry and a signal-processing framework capable of extracting weak target echoes from the received data \cite{willis2005bistatic}. { The bistatic geometry considered in this work is illustrated in Fig.~\ref{fig:bistatic}. The direct-path signal propagates from the illuminator to the InteRFeX receiver over the baseline $R_{\mathrm{IR}}$, whereas the target echo follows the illuminator--target and target--receiver paths, denoted by $R_{\mathrm{IT}}$ and $R_{\mathrm{TR}}$, respectively. The corresponding bistatic excess path length is given by
\begin{equation}
r_b(t)=R_{\mathrm{IT}}(t)+R_{\mathrm{TR}}(t)-R_{\mathrm{IR}}.
\label{eq:bistatic_range}
\end{equation}
The signal-processing chain begins with the hybrid analogue--digital beamforming architecture described in Section~\ref{system}. The fixed zenith-pointed analogue beam of each $4\times4$ Tile defines the instantaneous field of view, while the 16 digitized Tile outputs are coherently combined in the digital domain to form simultaneous reference and surveillance beams, as shown in Fig.~\ref{fig:bistatic}. The reference beam uses a fixed digital steering vector corresponding to the known illuminator location, whereas the surveillance beam is steered toward the TLE-predicted target direction. Its steering vector is updated at each coherent processing interval (CPI) to follow the expected target trajectory throughout the observation pass.}

\begin{figure}[!t]
    \centering
    \includegraphics[width=\linewidth]{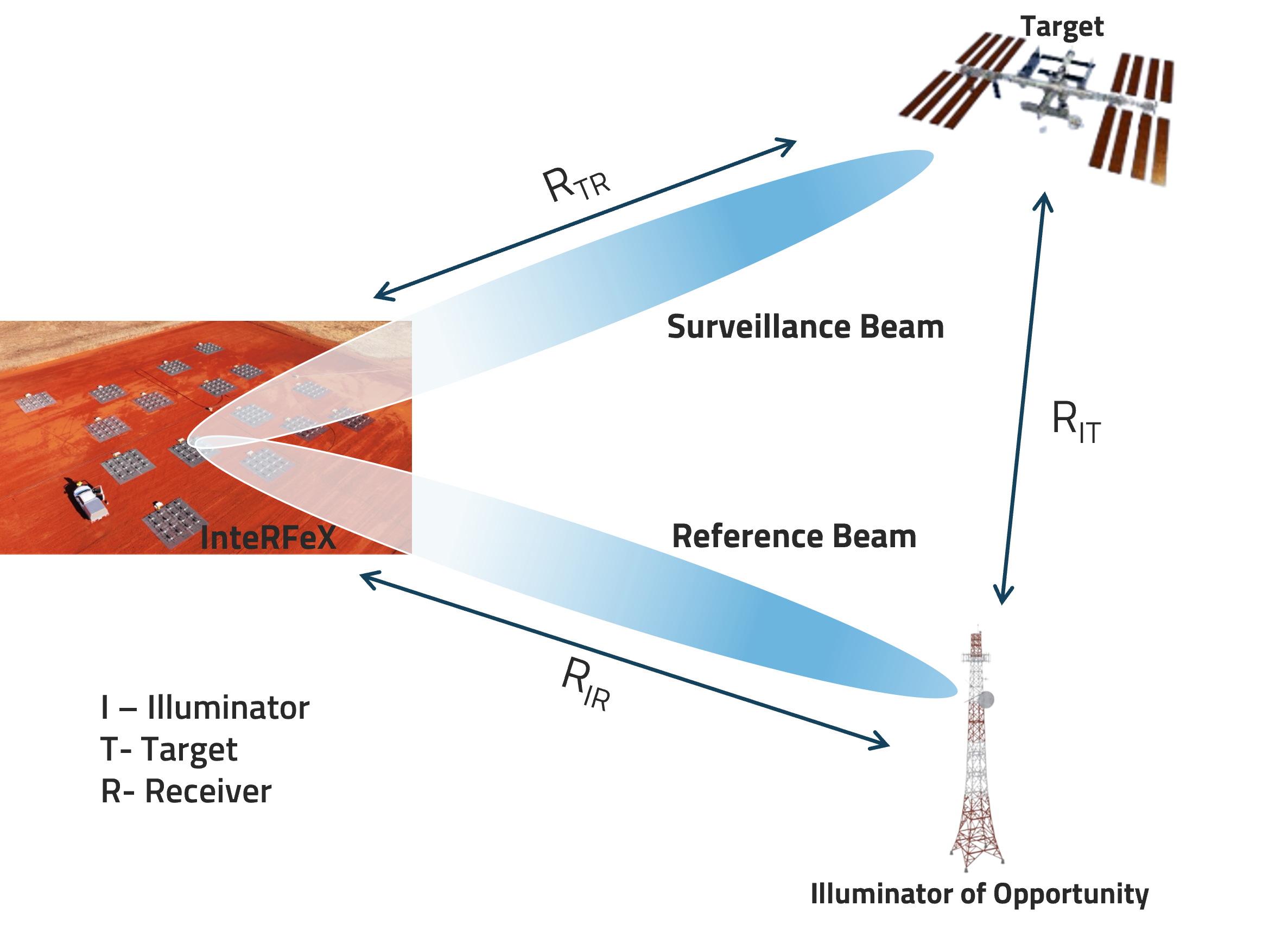}
    \caption{InteRFeX bistatic radar configuration for ISS observation, showing the illuminator of opportunity, receiver array, target, and reference and surveillance paths.}
    \label{fig:bistatic}
\end{figure}

{The illuminator of opportunity (in our case, an FM transmitter) emits a complex baseband waveform, denoted by $x(t)$. Since only the relative delay and phase evolution are relevant for subsequent correlation processing, constant amplitude attenuation and propagation phase terms are omitted, while the propagation delay and range-dependent carrier phase are explicitly preserved \cite{malanowski2019signal}. As the waveform propagates through the environment, it reaches the receiver via two primary paths: a direct path from the transmitter and a reflected path from the target. Based on the bistatic geometry illustrated in Fig.~\ref{fig:bistatic}, the reference signal corresponding to the direct-path transmission, denoted by $x_{ref}(t)$, is given by}

\begin{equation}
    x_{ref}(t) = x\left(t-\frac{R_{IR}}{c}\right).
\end{equation}
{Since the direct-path distance remains approximately constant throughout the observation interval, the geometric component of the propagation delay is treated as constant. Accordingly, its associated phase contribution is absorbed into the reference-channel calibration. However, this approximation does not account for time-varying propagation effects arising from atmospheric or environmental conditions, which may introduce additional phase variations, particularly for low-elevation or below-horizon propagation paths \cite{barott}. Such time-varying propagation effects are not explicitly modelled in the present analysis. The surveillance beam, on the other hand, receives the target echo, which propagates along the bistatic path comprising the illuminator-to-target and target-to-receiver links. Accordingly, the echo signal, $x_{echo}(t)$ is modelled as}

\begin{equation}
\begin{aligned}
    x_{echo}(t) =\;& x\!\left(t-\frac{R_{IT}(t)+R_{TR}(t)}{c}\right) \\
    &\times \exp\!\left(-j\,\frac{2\pi\left(R_{IT}(t)+R_{TR}(t)\right)}{\lambda}\right).
\end{aligned}
\end{equation}
{Since $x(t)$ represents the complex baseband waveform, the time-delay term models the propagation of the baseband envelope, whereas the exponential term represents the propagation-induced carrier phase associated with the varying bistatic path length.}

\begin{figure*}[!t]
    \centering
    \includegraphics[width=\textwidth]{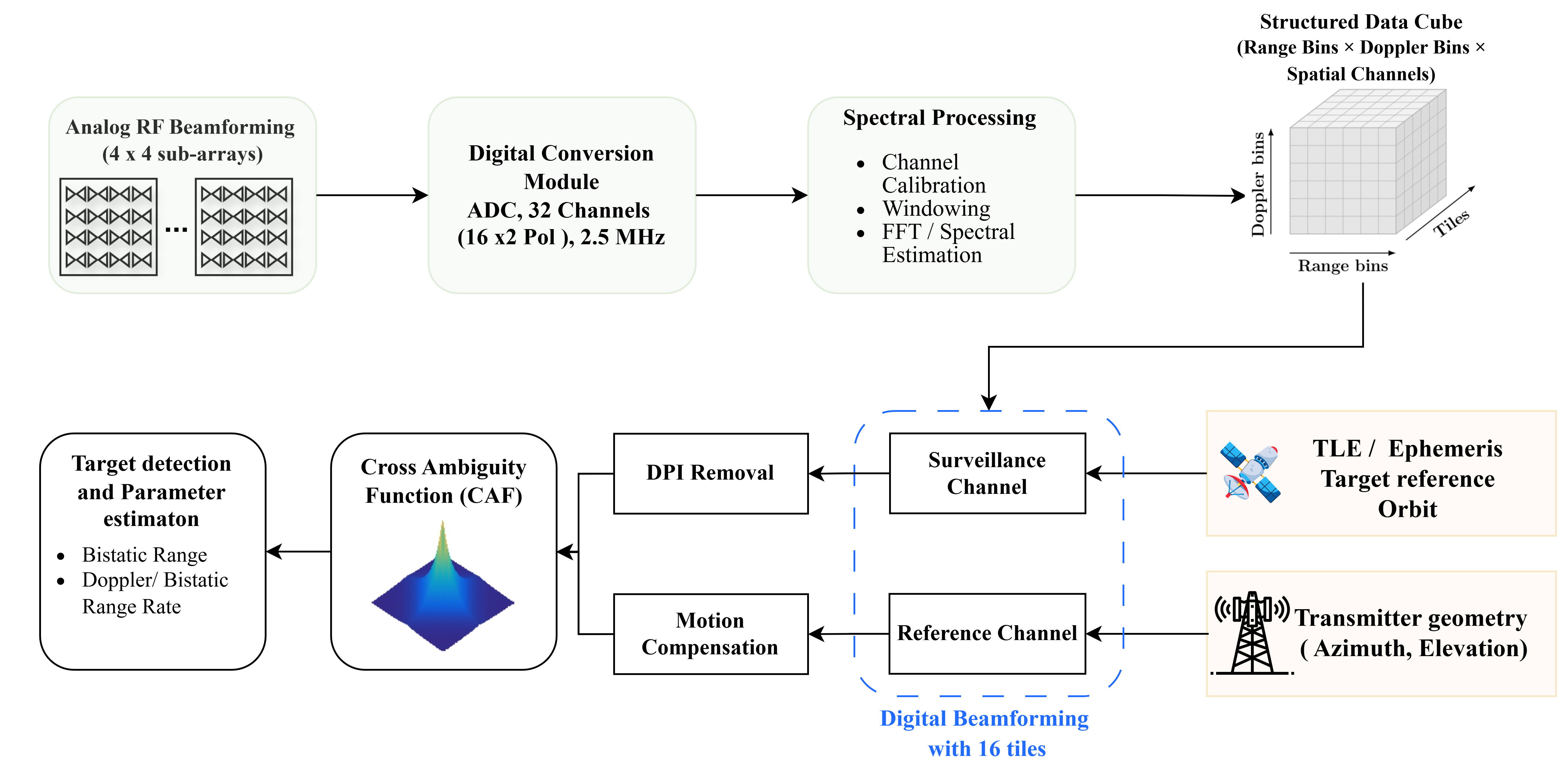}
    \caption{Flowchart of the proposed signal processing framework, illustrating the complete processing chain from beamforming to target parameter estimation.}
    \label{flowchart}
\end{figure*} 

Following the formation of the surveillance and reference beams, the surveillance channel typically contains significant clutter and direct-path interference (DPI) from the illuminator. Since the target echo is considerably weaker than these interference components, a clutter suppression technique, such as an adaptive filter, is applied prior to target detection \cite{malanowski2019signal}.  Once the clutter is mitigated, the target trajectory can be characterized over a specified observation interval. It is assumed that the target remains within the main lobe ($3~\mathrm{dB}$ beamwidth) throughout a single CPI, typically on the order of $0.5$--$1~\mathrm{s}$ in passive radar applications. 
{ 
The target motion over this interval is characterized by the instantaneous bistatic excess path length, $r_b(t)$, given in \eqref{eq:bistatic_range}.
Over a single CPI, the target motion is modelled by a Taylor expansion of the instantaneous bistatic excess path length, $r_b(t)$, about the midpoint of the CPI, denoted by $t_m$,

\begin{equation}
r_b(t)
=
r_b(t_m)
+
\left.\frac{dr_b(t)}{dt}\right|_{t=t_m}(t-t_m)
+
\mathcal{O}\!\left((t-t_m)^2\right).
\end{equation}
Since the higher-order motion terms are assumed to have a negligible influence within a CPI, the expansion is truncated after the first-order term. Introducing the local time variable $\tau=t-t_m$, such that $\tau=0$ corresponds to the CPI midpoint, gives

\begin{equation}\label{birange}
r_b(\tau)\approx R+V\tau,
\end{equation}
where $R=r_b(t_m)$ is the bistatic excess path length evaluated at the CPI midpoint, and
\begin{equation}
V=\left.\frac{dr_b(t)}{dt}\right|_{t=t_m}
\end{equation}
is the rate of change of the bistatic excess path length with respect to time, evaluated at the CPI midpoint. The objective of the detection stage is to estimate these bistatic parameters from the received echo using the Cross Ambiguity Function (CAF), which performs a coherent correlation between the surveillance and reference channel \cite{malanowski2019signal}. For LEO targets, however, an exhaustive search over the entire range--Doppler space is computationally demanding due to the large search region. To reduce this computational burden, prior orbital information is incorporated into the signal processing chain. Specifically, publicly available TLE data\footnote{TLE data was obtained from Space-Track.org, \url{https://www.space-track.org}} are propagated to the measurement epoch using the SGP4 propagator \cite{vallado2008sgp4} to predict the nominal bistatic excess path trajectory, denoted by $\tilde r_b(\tau)$. This predicted trajectory is used to motion-compensate the reference signal by applying the corresponding propagation delay and range-dependent carrier phase,

\begin{equation}\label{shift}
x_{ref}^{comp}(\tau)
=
x_{ref}\!\left(
\tau-\frac{\tilde r_b(\tau)}{c}
\right)
\exp\!\left(
-j\frac{2\pi\tilde r_b(\tau)}{\lambda}
\right).
\end{equation}
The compensated reference signal therefore represents the echo expected from the TLE-predicted trajectory. {Consequently, the CAF only needs to estimate the residual delay--Doppler mismatch relative to the TLE-derived reference.} The residual bistatic range ($\Delta R$) and bistatic range rate ($\Delta V$) are defined as

\begin{equation}
\Delta R = R-\tilde{R},
\hspace{1.0cm}\Delta V = V-\tilde{V},
\end{equation}
where $\tilde{R}$ and $\tilde{V}$ denote the TLE-predicted bistatic excess path length and bistatic range rate, respectively. The CAF is then computed between the surveillance signal and the motion-compensated reference signal as

\begin{equation}\label{caf}
\begin{aligned}
\chi(\Delta R,\Delta V)
&=
\int_{-T/2}^{T/2}
x_{echo}(\tau)\,
x_{ref}^{comp*}
\!\left(
\tau-\frac{\Delta R}{c}
\right) \\
&\qquad\times
\exp\!\left(
+j\frac{2\pi}{\lambda}\Delta V\,\tau
\right)
\,d\tau,
\end{aligned}
\end{equation}
where $T$ denotes the CPI and $(\cdot)^*$ represents complex conjugation. A correlation peak is obtained when the residual delay and phase compensation match the true target motion, thereby estimating the deviations from the TLE prediction. {The estimated residuals can then be combined with the TLE-predicted bistatic parameters to obtain corrected estimates of the target bistatic range and range rate.} The complete signal-processing framework described in this section is summarized in Fig.~\ref{flowchart}.

In the proposed framework, the TLE information is used solely to initialize the motion compensation and constrain the subsequent CAF search to a local residual parameter space. Unlike pulsed radar, passive radar employing continuous-wave broadcast illuminators does not generally suffer from the classical range and Doppler ambiguities associated with pulse repetition frequency, except in certain cases where the repetitive structure of the transmitted waveform introduces measurement ambiguities \cite{malanowski2019signal}. Moreover, following motion compensation, the expected target response is centered near zero residual delay and Doppler, allowing the CAF search to be confined to a local residual delay–Doppler region surrounding the predicted trajectory. This suppresses spurious competing CAF peaks, thereby reducing the search space and significantly lowering the computational burden. 

The underlying CAF formulation, however, is not inherently dependent on prior orbital information. In the absence of prior orbital information, such as for uncatalogued space debris, the same processing framework remains applicable. Alternative search strategies, including Orbit Determination Before Detect (ODBD) and higher-order bistatic motion models, have been reported for detection without prior orbital information. These approaches typically require searching a much larger bistatic parameter space, resulting in significantly increased computational complexity \cite{jkedrzejewski2025search,hennessy2020orbit}.}

\section{Experimental Results and Discussion}\label{results}

\begin{figure}[!t]
    \centering
    \includegraphics[width=.95\linewidth]{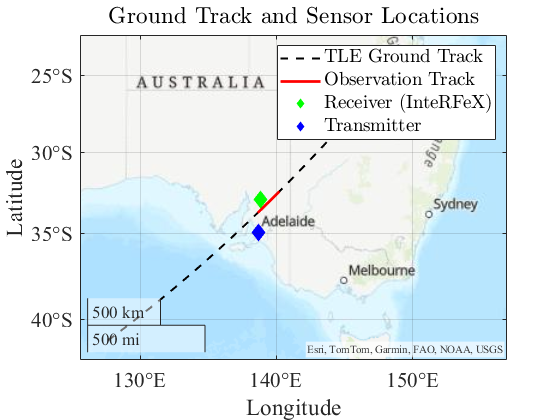}
    \caption{Ground track of the target over the transmitter–receiver region during the observation window.}
    \label{gt}
\end{figure}

An initial measurement campaign was conducted to assess the detection capability of InteRFeX, with the ISS selected as the target due to its large radar cross section (RCS). Data acquisition commenced on 24~November~2025 at 23:32:45~UTC, with a center frequency of $107.9~\mathrm{MHz}$. Owing to the maximum available FM transmitter bandwidth of approximately $200~\mathrm{kHz}$, the achievable range resolution is relatively coarse. In comparison, fine Doppler resolution can be achieved by appropriately selecting the CPI \cite{malanowski2019signal}

\begin{figure}[!t]
\centering

\begin{subfigure}[b]{0.49\linewidth}
\centering
\includegraphics[width=\linewidth]{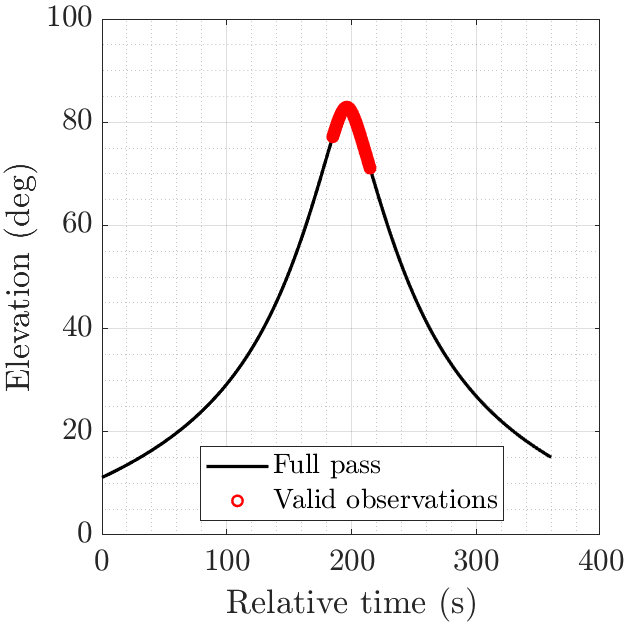}
\caption{}
\end{subfigure}
\begin{subfigure}[b]{0.49\linewidth}
\centering
\includegraphics[width=\linewidth]{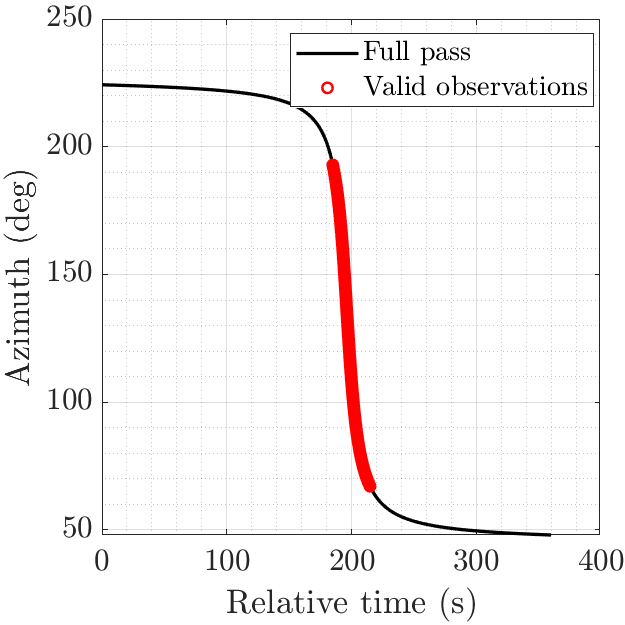}
\caption{}
\end{subfigure}

\begin{subfigure}[b]{0.49\linewidth}
\centering
\includegraphics[width=\linewidth]{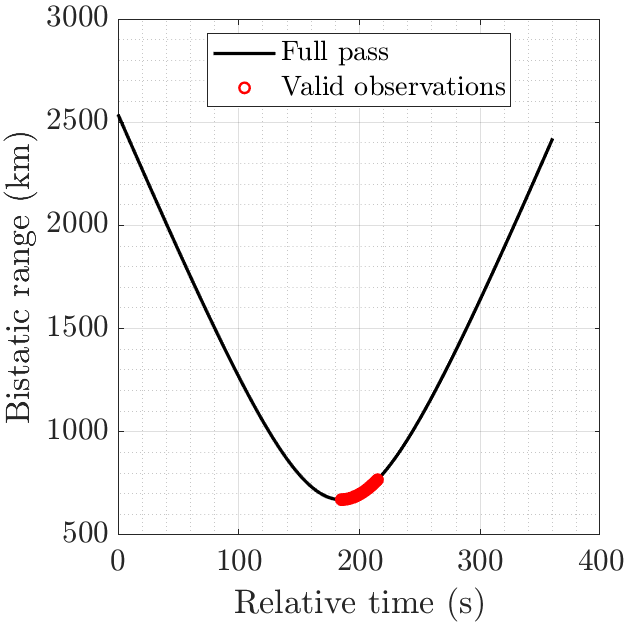}
\caption{}
\end{subfigure}
\begin{subfigure}[b]{0.49\linewidth}
\centering
\includegraphics[width=\linewidth]{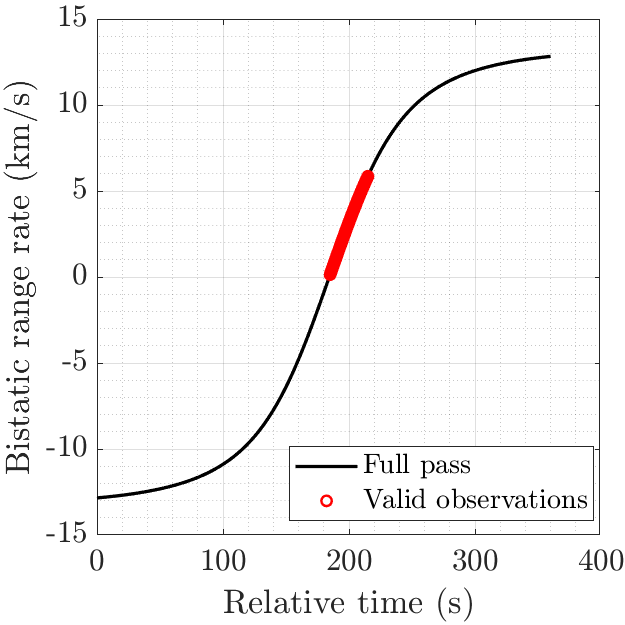}
\caption{}
\end{subfigure}

\caption{TLE-derived time histories of (a) topocentric elevation, (b) topocentric azimuth, (c) bistatic excess path length, and (d) bistatic range rate over the reference pass. The black curves represent the complete predicted pass, while the red markers indicate the observation samples used in the analysis.}
\label{fig:meas_time_series}
\end{figure}

For this experiment, an FM transmitter, 5RAM (Mt. Lofty), used as an IoO, is located in the Adelaide region ($34.978^\circ$ S, $138.708^\circ$ E), approximately $220~\mathrm{km}$ from the receiver at an azimuth of $183^\circ$, with a maximum effective radiated power (ERP) of $20~\mathrm{kW}$ as report by Australian Communications and Media Authority (ACMA)\cite{ACMA_Broadcast_Transmitters}. Fig.~\ref{gt} illustrates the ISS ground track over South Australia around the measurement time, obtained by propagating its orbit using TLE data. The InteRFeX site and transmitter location are indicated by green and blue diamonds, respectively, while the dashed black line represents the propagated ISS ground track.
A 30-s interval, highlighted by the red colour in Fig.~\ref{gt}, was selected for detailed analysis, as it corresponds to the period during which the ISS remained within the Tile field of view and yielded consistent detections. The elevation, azimuth, bistatic range, and bistatic range rate predicted over the pass by propagating the ISS orbit using TLE data are shown as functions of relative time in Fig.~\ref{fig:meas_time_series}. During the considered observation interval, the bistatic angle varies from approximately $24^\circ$ to $29^\circ$, corresponding to the region near the maximum bistatic angle attained during this pass. The changing bistatic geometry can influence the effective target RCS and, consequently, the received power and SNR. The following subsection presents the link-budget and predicted SNR analysis for the considered observation pass. Moreover, TLE information was used to steer the surveillance beam and compensate for target motion, while angular estimation was performed independently using the received data. The resulting angular estimates were used to validate the observed trajectory and assess the angular resolution capability of InteRFeX. The experimental results are presented in the following subsections (Section \ref{sec:doa} $\&$ \ref{sec:caf}).

{

\subsection{Link-Budget Analysis and SNR Prediction}
\label{subsec:link_budget}

\begin{table}[t]
\caption{{Parameters Used for the Bistatic SNR Prediction.}}
\label{tab:link_budget_parameters}
\centering
\footnotesize
\renewcommand{\arraystretch}{1.08}

\begin{tabularx}{\columnwidth}{Xcc}
\hline
\textbf{Parameter} & \textbf{Symbol} & \textbf{Value} \\
\hline

Maximum illuminator EIRP
& $\mathrm{EIRP}_{\mathrm{I,max}}$
& $32.8$~kW \\

Effective illuminator EIRP
& $\mathrm{EIRP}_{\mathrm{I,eff}}$
& $3.28$~kW \\

Operating frequency
& $f_0$
& $107.9$~MHz \\

Wavelength
& $\lambda$
& $2.78$~m \\

Processed bandwidth
& $B$
& $150$~kHz \\

Coherent processing interval
& $T_{\mathrm{CPI}}$
& $1$~s \\

Peak receive-array gain
& $G_{r,\max}$
& $\approx 27.03$~dBi \\

System noise temperature
& $T_{\mathrm{sys}}$
& $1000$~K \\

Aggregate system losses
& $L_{\mathrm{sys}}$
& $6$~dB \\

Bistatic RCS
& $\sigma_{\mathrm{b}}$
& $400$~m$^2$ \\

\hline
\end{tabularx}

\end{table}

A first-order bistatic link-budget analysis is performed to assess the expected target detectability of the InteRFeX system. Assuming ideal coherent integration over a coherent processing interval $T_{\mathrm{CPI}}$, the predicted output SNR \cite{malanowski2019signal} is expressed as
\begin{equation}
\mathrm{SNR(t)}
=
\frac{
\mathrm{EIRP}_{\mathrm{I,eff}}\,
G_r(\theta,\phi)\,
\sigma_{\mathrm{b}}\,
\lambda^2\,
T_{\mathrm{CPI}}
}{
(4\pi)^3
R_{\mathrm{IT}}^2(t)\,
R_{\mathrm{TR}}^2(t)\,
k_BT_{\mathrm{sys}}L_{\mathrm{sys}}
},
\label{eq:predicted_snr}
\end{equation}
where $G_r(\theta,\phi)$ is the receive-array gain in the instantaneous target direction, $\sigma_{\mathrm{b}}$ is the effective bistatic RCS, and $R_{\mathrm{IT}}(t)$ and $R_{\mathrm{TR}}(t)$ are the time-varying distances from the illuminator to the target and from the target to the receiver, respectively, as illustrated in Fig.~\ref{fig:bistatic}. The terms $T_{\mathrm{sys}}$ and $L_{\mathrm{sys}}$ denote the effective system noise temperature and aggregate implementation losses, respectively, with $L_{\mathrm{sys}}$ accounting for losses in the analogue receiver chain, cables, beamforming, calibration, and signal processing. The quantity effective isotropic radiated power, $\mathrm{EIRP}_{\mathrm{I,eff}}$ in \eqref{eq:predicted_snr} denotes the effective isotropic radiated power toward the target. For the $107.9~\mathrm{MHz}$ 5RAM FM illuminator, the ACMA database specifies a maximum ERP of $20~\mathrm{kW}$~\cite{ACMA_Broadcast_Transmitters}. Since ERP is referenced to a half-wave dipole, whereas the link-budget formulation uses an isotropic reference, it is converted using $\mathrm{EIRP}=1.64 \times \mathrm{ERP}$~\cite{saunders2024antennas}, yielding a maximum EIRP of approximately $32.8~\mathrm{kW}$. Additionally, the transmitter radiation pattern toward the target is unavailable; a conservative directional attenuation of $10~\mathrm{dB}$ is applied. Consequently, the effective EIRP used in \eqref{eq:predicted_snr} is
\begin{equation}
\mathrm{EIRP}_{\mathrm{I,eff}}
=
\mathrm{EIRP}_{\mathrm{I,max}}10^{-10/10}
\approx 3.28~\mathrm{kW}.
\end{equation}
Moreover, an effective bistatic RCS of $\sigma_{\mathrm{b}}=400~\mathrm{m}^{2}$ is assumed for the ISS in the link-budget analysis, consistent with nominal catalogue RCS values adopted in previous MWA and LOFAR passive-radar studies~\cite{prabu2020development,jkedrzejewski2025analysis}. This value is used only as a first-order approximation for the considered bistatic geometry. 

For the link-budget analysis, the receive-array gain, $G_r(\theta,\phi)$, was obtained from the simulated full-array radiation pattern based on the two-stage beamforming architecture described in Section~\ref{system}. The simulation accounts for the radiation pattern of the individual dipole elements, which forms the element-level response of the array. The analogue combination of the $4\times4$ elements within each Tile provides an ideal coherent gain of approximately $12~\mathrm{dB}$, while the subsequent digital combination of the $16$ Tile outputs provides an additional ideal coherent gain of approximately $12~\mathrm{dB}$. These element-pattern and beamforming contributions are incorporated into the simulated full-array radiation pattern. The resulting direction-dependent gain was evaluated along the predicted target trajectory and varies from approximately $24$ to $27~\mathrm{dB}$ over the observation pass. The array-pattern simulation neglects mutual coupling between the elements, and the resulting direction-dependent receive gain is subsequently used in the link-budget analysis. The remaining parameters adopted for the SNR prediction are summarized in Table~\ref{tab:link_budget_parameters}. Using these parameters, the predicted SNR over the complete ISS pass was calculated using \eqref{eq:predicted_snr} and is shown in Fig.~\ref{fig:predicted_snr}. To account for uncertainty in the transmitter gain toward the target, the transmitter-pattern loss is varied from $5$ to $15~\mathrm{dB}$, with $10~\mathrm{dB}$ adopted as the nominal case. The corresponding variation in the predicted SNR is represented by the shaded region.

\begin{figure}[t]
    \centering
    \includegraphics[width=\columnwidth]{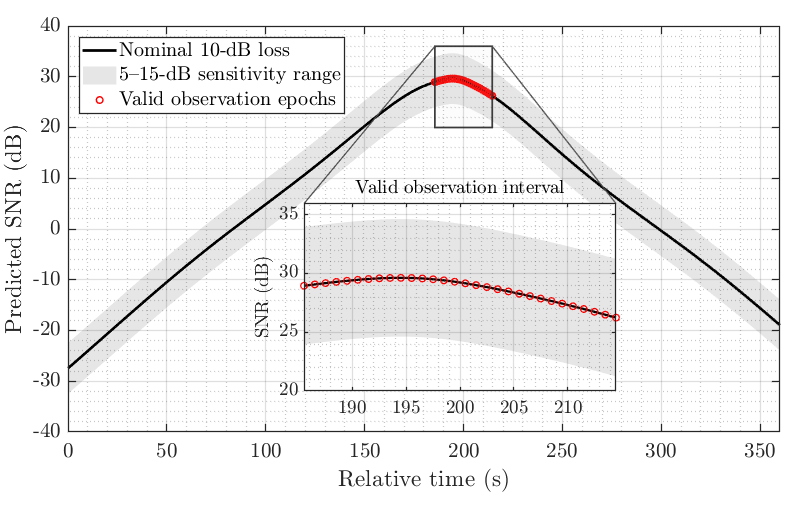}
\caption{{Predicted SNR of the ISS echo over the complete pass for a nominal transmitter-pattern loss of $10$~dB. The shaded region represents the range corresponding to transmitter-pattern losses of $5$--$15$~dB, and the inset enlarges the valid observation interval indicated by red circles.}}
    \label{fig:predicted_snr}
\end{figure}

The predicted SNR values corresponding to valid observations are indicated by red open circles and enlarged in the inset of Fig.~\ref{fig:predicted_snr}. During this interval, the nominal predicted SNR for the assumed $10\text{-}\mathrm{dB}$ transmitter-pattern loss decreases from approximately $29$ to $26~\mathrm{dB}$, while the $5$--$15~\mathrm{dB}$ sensitivity range remains approximately $\pm5~\mathrm{dB}$ relative to the nominal prediction. The temporal SNR variation is primarily governed by the time-varying illuminator--target and target--receiver distances, $R_{\mathrm{IT}}(t)$ and $R_{\mathrm{TR}}(t)$, respectively. Their combined variation produces a broad SNR maximum near the middle of the pass and lower values toward its beginning and end. The predicted values suggest favourable detectability of the ISS echo throughout the available observation interval. Nevertheless, the absolute SNR values should be regarded as first-order estimates because of uncertainties in the transmitter radiation pattern, effective bistatic RCS, system noise temperature, and unmodeled system losses. Moreover, the present analysis assumes a bistatic RCS of $400~\mathrm{m}^2$ for the ISS. For lower-RCS targets, the predicted SNR decreases accordingly; under the same geometry and system parameters, RCS values of $10~\mathrm{m}^2$ and $1~\mathrm{m}^2$ correspond to SNR reductions of approximately $16$ and $26~\mathrm{dB}$, respectively. These cases illustrate the expected SNR scaling for smaller LEO targets and should not be interpreted as representative of specific satellites or debris, since the bistatic RCS varies with target geometry, aspect, frequency, and polarization.

The temporal stability of the beamformed reference channel was also evaluated over the 30-s observation interval using the valid non-overlapping 1-s CPIs. The reference-channel SNR was estimated from the power ratio between the $107.9~\mathrm{MHz}$ signal band and an adjacent signal-free band at $108.2~\mathrm{MHz}$, yielding mean $\pm$ standard deviation values of $16.20 \pm 0.15~\mathrm{dB}$ and $16.12 \pm 0.26~\mathrm{dB}$ for the X- and Y-polarized beams, respectively. The comparable SNR levels and limited temporal variation indicate that both polarizations provide stable, high-quality reference signals for the subsequent CAF processing discussed in Section~\ref{sec:caf}.
}
\subsection{Direction-of-Arrival Analysis}\label{sec:doa}

Several approaches exist for direction-of-arrival (DoA) estimation in  the passive radar domain, including delay-and-sum beamforming, minimum variance distortionless response (MVDR), and Multiple Signal Classification (MUSIC), each offering a different trade-off between resolution, computational complexity, and robustness \cite{jose_beamforming}. In this study, MUSIC is adopted due to its high angular resolution and its capability to resolve closely spaced sources, which has also been demonstrated to be effective for space target detection in systems such as BIRALES \cite{montaruli2022adaptive}. { DoA estimation was performed using a frequency-domain implementation of the MUSIC algorithm. The array voltage data were first divided into CPIs and transformed into frequency domain using block FFT processing. Prior to DoA estimation, frequency dependent complex gain calibration coefficients were applied to compensate for the channel-to-channel amplitude and phase mismatches. For each CPI, the frequency bin centered at $107.9~\mathrm{MHz}$ was selected for DoA estimation. The calibrated array snapshots for the selected polarization were used to estimate the sample spatial covariance matrix as
\begin{equation}
\widehat{\mathbf{R}}=\frac{1}{N}\sum_{n=1}^{N}
\mathbf{x}(n)\mathbf{x}^{H}(n),
\end{equation}
where $\mathbf{x}(n)$ denotes the complex array snapshot at $107.9~\mathrm{MHz}$ and $N$ is the number of snapshots within the CPI. An eigenvalue decomposition of the sample covariance matrix was subsequently performed. The signal subspace was constructed from the dominant eigenvectors associated with the largest eigenvalues, while the remaining eigenvectors defined the noise subspace. In the present implementation, four dominant eigenvectors were retained, as this configuration provided the most stable DoA estimates across the experimental dataset. The MUSIC pseudo-spectrum was evaluated by projecting the array steering vectors onto the estimated noise subspace. Using the measured antenna coordinates and selected RF frequency, the steering vectors were formulated in the direction-cosine domain defined in \eqref{eq:direction_cosines}. The MUSIC pseudo-spectrum was then computed as
\begin{equation}
P_{\mathrm{MUSIC}}(u,v)=
\frac{1}
{\mathbf{a}^{H}(u,v)\mathbf{E}_{n}\mathbf{E}_{n}^{H}\mathbf{a}(u,v)} \hspace{6pt} ,
\end{equation}
where $\mathbf{a}(u,v)$ denotes the array steering vector and $\mathbf{E}_{n}$ is the matrix containing the eigenvectors that span the estimated noise subspace. The resulting pseudo-spectrum was evaluated on a uniformly sampled direction-cosine grid over the visible hemisphere and normalized to its peak value prior to visualization, as shown in Fig.~\ref{doa_uv_map}. This DoA power spectrum was computed over 1-s CPI for a snapshot recorded on 24 November 2025 at 23:43:19 UTC using the X-polarization channel
}
\begin{figure}[t]
    \centering
    \includegraphics[width=0.95\linewidth]{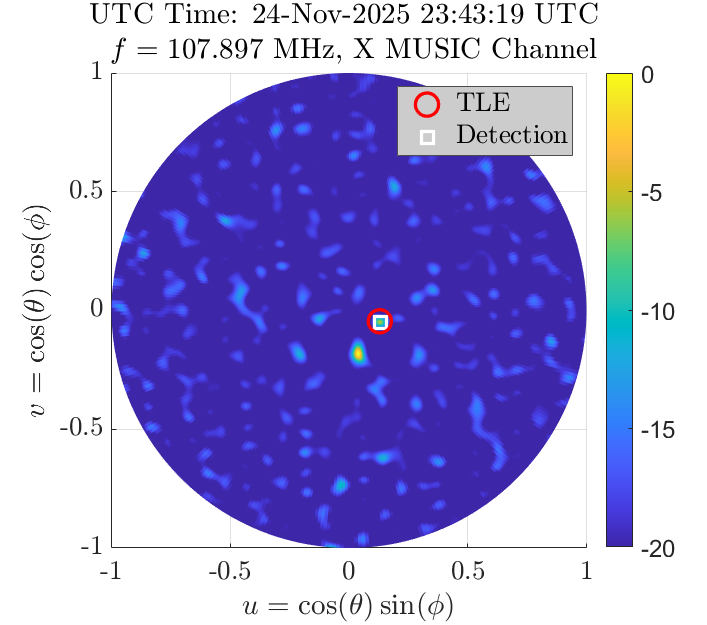}
    \caption{{MUSIC direction-of-arrival pseudo-spectrum for the X-polarized channel. The dominant peak corresponds to the ISS, while the persistent source below and to the left of the ISS is the astronomical radio galaxy Centaurus A, as discussed in the text.}}
    \label{doa_uv_map}
\end{figure}

In Fig.~\ref{doa_uv_map}, the red circle marker indicates the TLE-predicted direction of arrival, propagated to the measurement epoch via SGP4 propagator and projected into the local topocentric frame of the passive radar array. The white square marker indicates the MUSIC peak corresponding to the estimated DoA of the ISS. The two markers show close spatial agreement within the direction cosine plane, confirming that the array successfully resolves the target against the background. The detection peak stands roughly $17~\mathrm{dB}$ above the surrounding noise floor confirming a clean localization result for this interval. 
{It is also worth noting that a bright, persistent peak is visible in the map at \(u=0.03, v=-0.18\). This detection is consistent with the radio galaxy Centaurus~A, which serves as an incidental confirmation that the array is coherently sensitive to astronomical sources at these frequencies.}

To track the ISS across the full observation pass, a local-maximum detection algorithm was applied to successive DoA frames. The angular separation $\Delta\psi$ between each DoA estimate and the corresponding TLE-derived reference direction is defined as
\begin{equation}
\label{ang_sep}
\begin{aligned}
\Delta\psi = \arccos\Bigl[
    &\sin(el_e)\sin(el_r) \\
    &+ \cos(el_e)\cos(el_r)\cos(az_e-az_r)
\Bigr],
\end{aligned}
\end{equation}
where $(az_e,el_e)$ are the estimated azimuth and elevation obtained from the MUSIC peak, and $(az_r,el_r)$ are the corresponding TLE-derived reference values.  This metric is preferred because it is coordinate-frame invariant and provides a physically meaningful scalar measure of the angular discrepancy between the estimated and TLE-derived directions. {For clarity, all reported range, range-rate, and angular deviations throughout this paper are computed relative to the TLE-derived reference trajectory and should not be interpreted as absolute measurement errors.}

\begin{figure}[!t]
    \centering
    \includegraphics[width=0.98\linewidth]{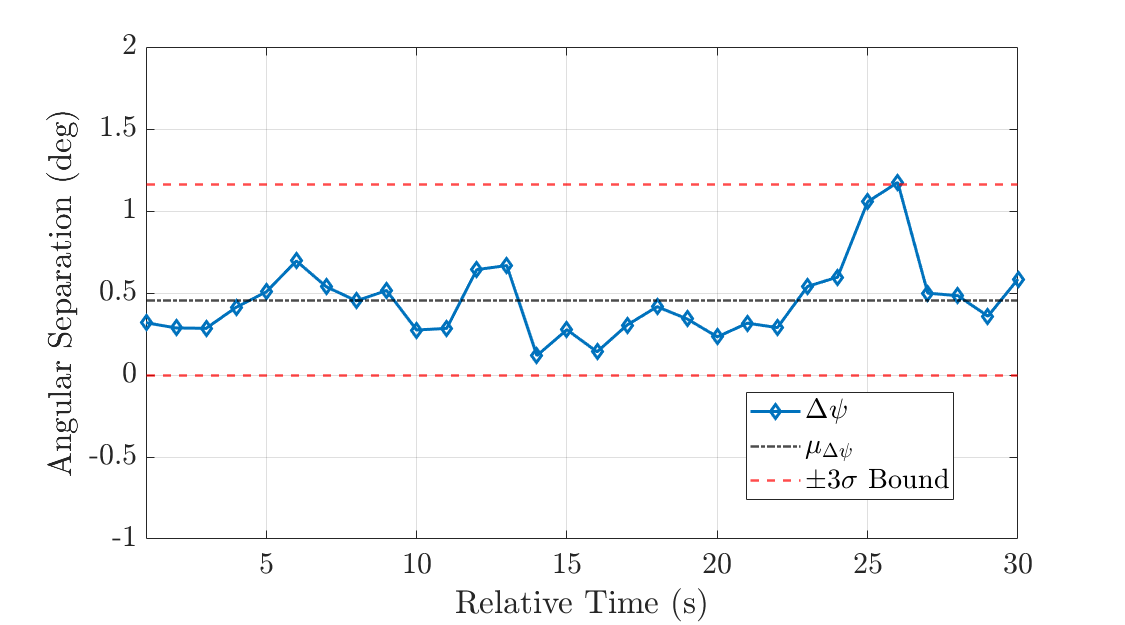}
    \caption{Angular separation $\Delta\psi$ between the estimated DoA and the TLE-predicted direction over the observation interval. The mean (black dashed) and $\pm3\sigma$ bounds (red dashed) are indicated.}
    \label{doa_uv_sep}
\end{figure}

Fig.~\ref{doa_uv_sep} shows the angular separation $\Delta\psi$ as a function of relative time, where each point corresponds to a single DoA estimate extracted from a successive frame. The ISS was detected in all 30 frames within the 30-second observation window. The mean angular separation $\mu_{\Delta\psi}$ across the pass is approximately $0.45^\circ$, indicated by the black dashed line, and the $ \pm 3\sigma$ bounds are shown as red dashed lines, where $\sigma$ denotes the standard deviation of $\Delta\psi$ over all detected frames. {The angular separation $\Delta\psi$ remains within the $\pm3\sigma$ bounds throughout the observation interval, except for a single frame at $t=26~\mathrm{s}$.} Moreover, the along-track and cross-track angular offsets, $\Delta\psi_\mathrm{AT}$ and $\Delta\psi_\mathrm{CT}$, are computed at each epoch by resolving the great-circle displacement from the TLE reference direction to the estimated DoA in a local track-aligned frame on the unit sphere \cite{benedikter2026tracer}. { The along-track axis is tangent to the TLE trajectory in the instantaneous direction of motion, whereas the cross-track axis is orthogonal to it in the local tangent plane. Accordingly, $\Delta\psi_\mathrm{AT}>0$ ($<0$) indicates that the estimate lies ahead of (behind) the predicted direction, while $\Delta\psi_\mathrm{CT}>0$ ($<0$) indicates that it lies to the left (right) of the predicted trajectory.

\begin{figure}[!t]
    \centering
    \begin{subfigure}[b]{\linewidth}
        \includegraphics[width=\linewidth]{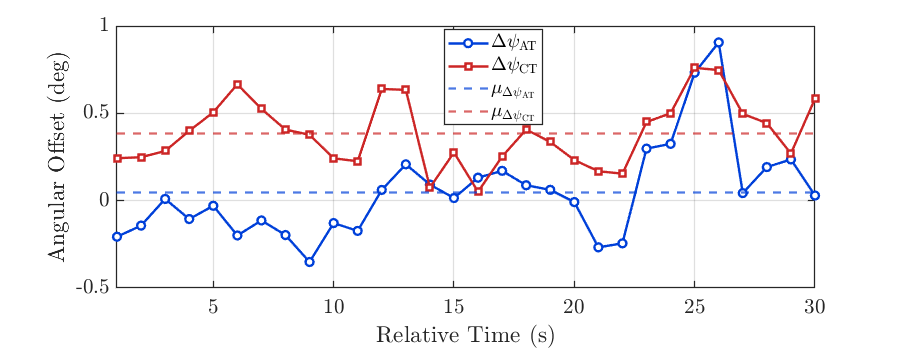}
        \caption{}
        \label{fig:Doa_at_ct_error}
    \end{subfigure}
   \hfill
    \begin{subfigure}[b]{\linewidth}
        \includegraphics[width=\linewidth]{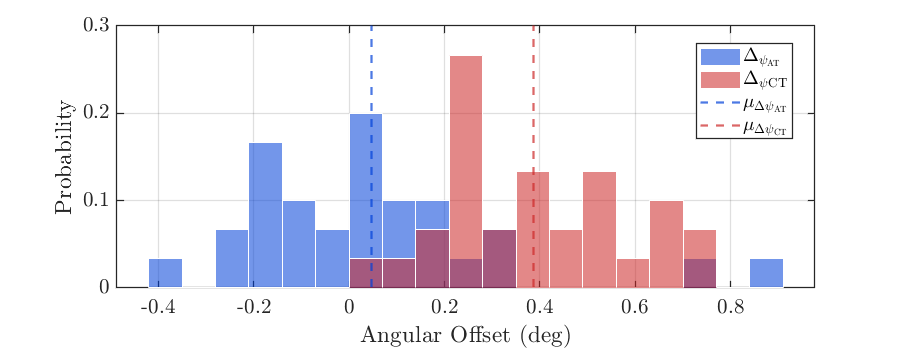}
        \caption{}
        \label{fig:doa_at_ct_histogram}
    \end{subfigure}
\caption{Along- and cross-track angular offsets: (a) temporal variation over the observation interval, with dashed lines indicating the respective mean values; and (b) corresponding probability distributions showing the spread about each mean.
}
    \label{fig:error_doa}
\end{figure}
Fig.~\ref{fig:Doa_at_ct_error} shows the temporal evolution of the along-track and cross-track angular offsets over the 30-s observation interval, with the dashed lines indicating their respective mean values. The along-track component fluctuates about a value close to zero, whereas the cross-track component remains predominantly positive throughout the pass, indicating a persistent transverse displacement relative to the TLE-predicted trajectory. The corresponding histograms in Fig.~\ref{fig:doa_at_ct_histogram} further highlight this difference: the along-track distribution is centred near zero, while the cross-track distribution is shifted toward positive offsets and exhibits a larger mean. This separation suggests that the dominant systematic component of the angular residual is cross-track rather than along-track. As $\Delta\psi_{\mathrm{CT}}$ measures displacement perpendicular to the satellite track, a timing misalignment would be expected to affect primarily the along-track component $\Delta\psi_{\mathrm{AT}}$, rather than produce a systematic cross-track offset. The observed cross-track bias may therefore result from residual phase-calibration errors or a small misalignment in the array pointing model. Multidimensional subspace-based DoA estimators are known to be sensitive to array-manifold and gain--phase errors \cite{swindlehurst1992performance}. To examine whether residual phase errors could produce this signature, a controlled phase-gradient sensitivity analysis was performed. Specifically, linear phase gradients of $\pm0.005~\mathrm{rad/m}$ were applied to the calibrated complex array data at the frequency bin selected for MUSIC processing, prior to covariance-matrix estimation. The gradients were oriented at $0^\circ$, $45^\circ$, $90^\circ$, and $135^\circ$, while the nominal MUSIC steering manifold was retained. 

\begin{table}[t]
    \centering
    \caption{{Phase-gradient sensitivity of the along-track (AT) and cross-track (CT) angular offsets.}}
    \label{tab:phase_gradient_sensitivity}
    \footnotesize
    \renewcommand{\arraystretch}{1.15}
    \begin{tabular*}{\columnwidth}
        {@{\extracolsep{\fill}}cccc@{}}
        \hline
        \shortstack{\textbf{Gradient}\\\textbf{orientation} ($^\circ$)} &
        \shortstack{\textbf{AT response}\\($^\circ$)} &
        \shortstack{\textbf{CT response}\\($^\circ$)} &
        \shortstack{\textbf{Dominant}\\\textbf{component}} \\
        \hline
        0   & 0.0179 & 0.0154 & Both        \\
        45  & 0.0575 & 0.0068 & Along-track \\
        90  & 0.0214 & 0.0180 & Both        \\
        135 & 0.0015 & 0.0820 & Cross-track \\
        \hline
    \end{tabular*}
\end{table}

Table~\ref{tab:phase_gradient_sensitivity} summarizes the response amplitudes, defined as one-half of the absolute difference between the changes produced by the positive and negative gradients relative to the zero-gradient case. The results show that the induced DoA displacement depends strongly on the gradient orientation, with some orientations producing mainly along-track shifts and others producing mainly cross-track shifts. The largest cross-track response was $0.0820^\circ$, corresponding to approximately $21\%$ of the measured cross-track bias. Residual phase-calibration errors may therefore contribute to the observed bias, while the remaining offset may arise from array-pointing misalignment and other systematic model errors. The mean and standard deviation of the along-track and cross-track offsets are summarized in Table~\ref{tab:doa_stats}. Although the along-track mean remains close to zero, its variability is higher, with $\sigma_{\Delta\psi_{\mathrm{AT}}}=0.27^\circ$. Overall, the decomposition shows that the dominant residual is cross-track, with residual phase-calibration errors representing one plausible contributor to the observed bias.}

\begin{figure*}[!tbp]
    \centering
    \begin{subfigure}[b]{0.49\linewidth}
        \includegraphics[width=\linewidth]{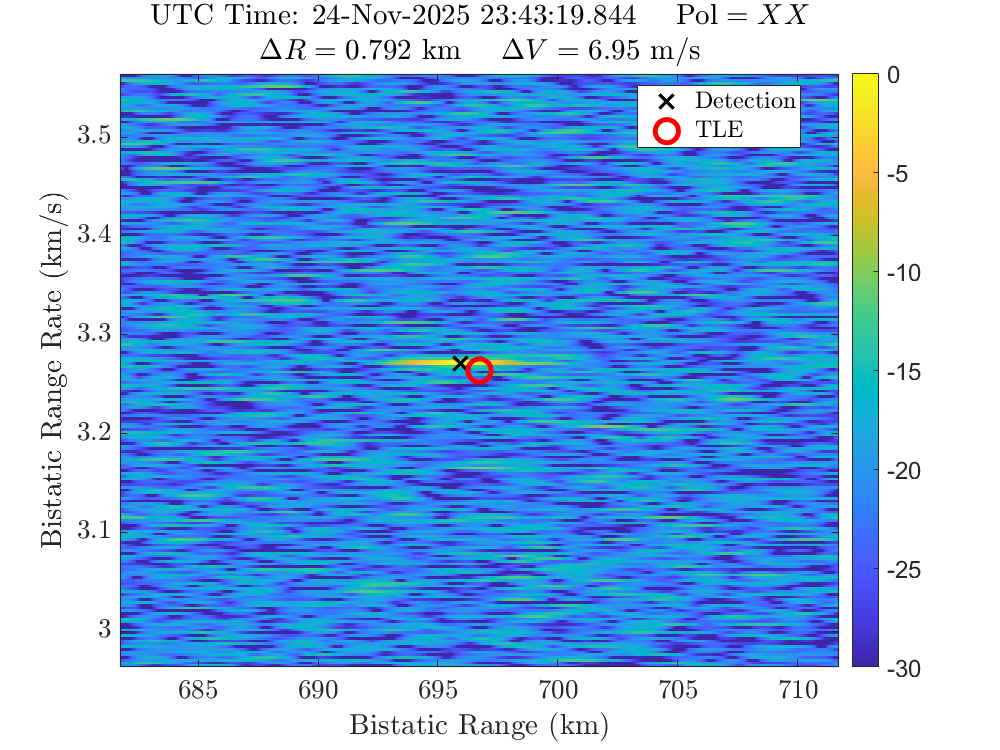}
        \caption{}
        \label{caf_XX}
    \end{subfigure}
    \hfill
    \begin{subfigure}[b]{0.49\linewidth}
        \includegraphics[width=\linewidth]{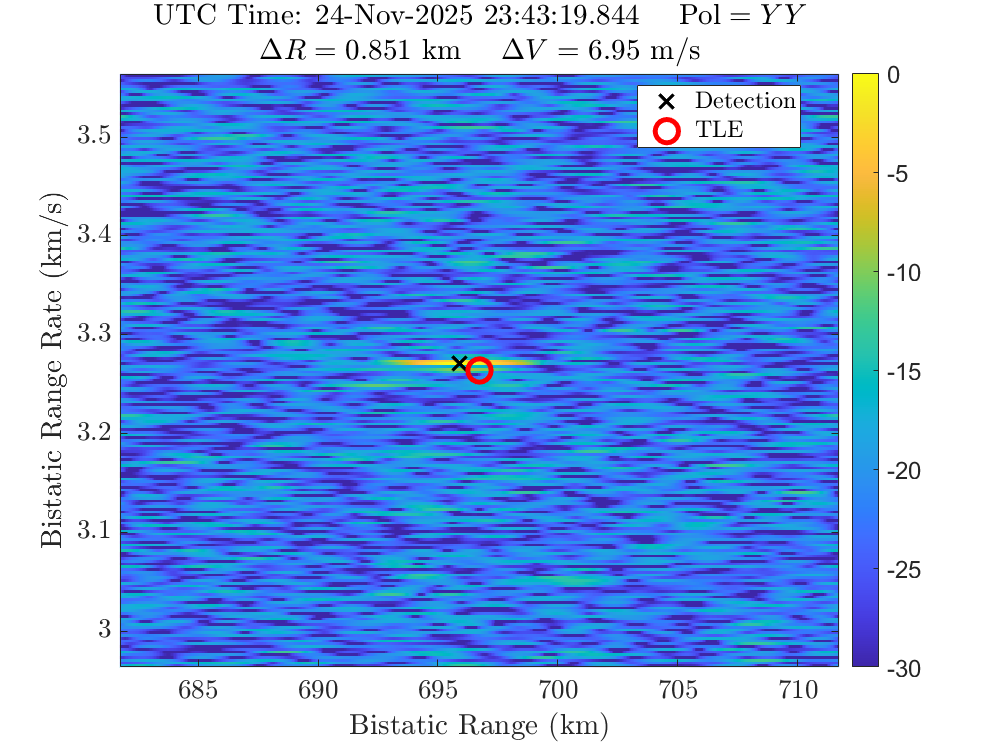}
        \caption{}
        \label{caf_YY}
    \end{subfigure}

    \vspace{0.5em}

    \begin{subfigure}[b]{0.49\linewidth}
        \includegraphics[width=\linewidth]{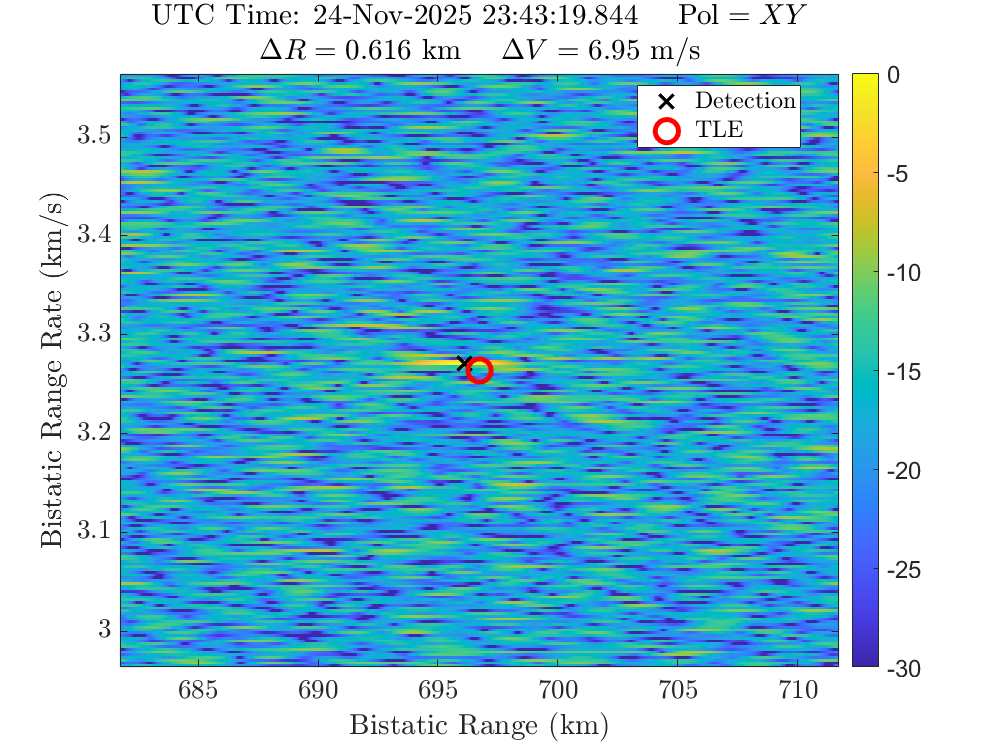}
        \caption{}
        \label{caf_XY}
    \end{subfigure}
    \hfill
    \begin{subfigure}[b]{0.49\linewidth}
        \includegraphics[width=\linewidth]{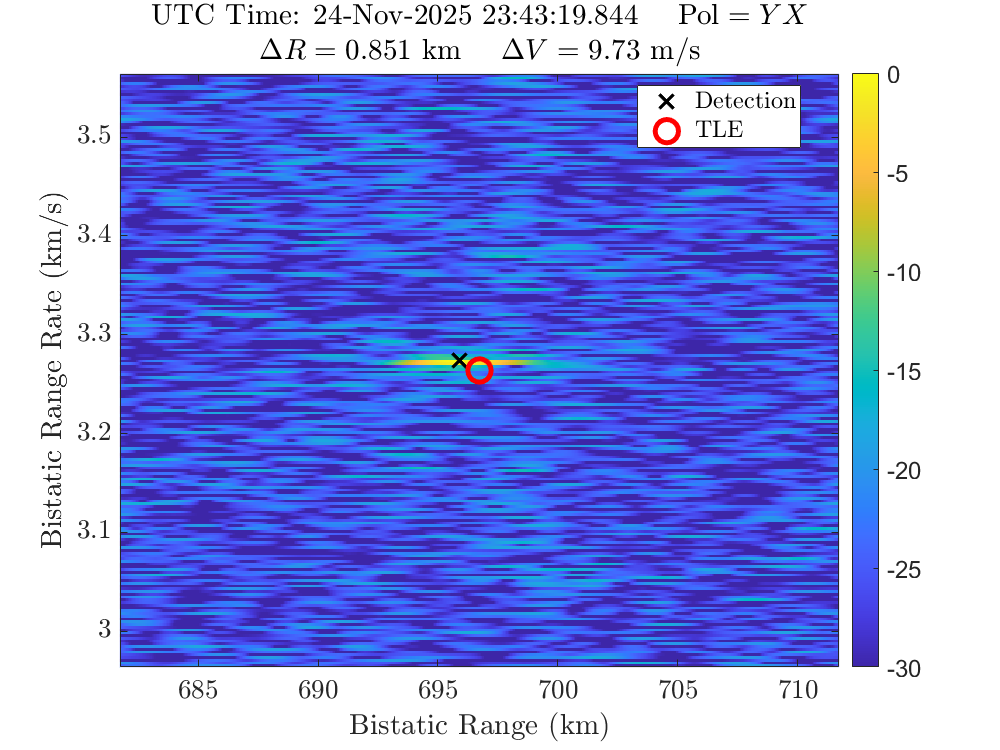}
        \caption{}
        \label{caf_YX}
    \end{subfigure}

    \caption{CAF zoom around the expected ISS bistatic range and range rate (red circle) for a selected timestamp, shown for different polarization combinations: (a) XX, (b) YY, (c) XY, and (d) YX.}
    \label{fig_caf}
\end{figure*}

\begin{table}[!tb]
\centering
\caption{Summary of the MUSIC DoA estimation performance metrics.}
\label{tab:doa_stats}
\renewcommand{\arraystretch}{1.3}

\begin{tabularx}{\columnwidth}{Xcc}
\hline
\textbf{Metric} & \textbf{\(\mu\)} & \textbf{\(\sigma\)} \\
\hline
Detection rate
& \multicolumn{2}{c}{100\%} \\
\hline
Angular separation \(\Delta\psi\)
& \(0.45^{\circ}\) & \(0.23^{\circ}\) \\

Along-track offset \(\Delta\psi_{\mathrm{AT}}\)
& \(+0.05^{\circ}\) & \(0.27^{\circ}\) \\

Cross-track offset \(\Delta\psi_{\mathrm{CT}}\)
& \(+0.38^{\circ}\) & \(0.19^{\circ}\) \\
\hline
\end{tabularx}

\end{table}

\subsection{Cross-Ambiguity Function Analysis}\label{sec:caf}

As described in Section~\ref{SP}, reference and surveillance beams were formed by compensating geometry-induced delays through phase adjustments, thereby increasing gain in the desired directions. Prior to correlation, the reference signal was compensated for the expected target delay and Doppler using \eqref{shift}, with the delay and Doppler parameters derived from the TLE-predicted trajectory. For each detected frame, the CAF was computed using~\eqref{caf}, yielding estimates of the bistatic range and range-rate deviations relative to the TLE-derived predictions. {To investigate the effect of polarization diversity on the CAF response, the CAF was evaluated for the four reference--surveillance combinations XX, XY, YX, and YY. These combinations are denoted by $pq$, where $p$ and $q$ identify the X/Y channels used to form the reference and surveillance signals, respectively. As defined in Section~\ref{system}, X and Y correspond to the East--West and North--South linear-polarization orientations of the array elements. Accordingly, XX and YY represent same-polarization-channel combinations, whereas XY and YX represent orthogonal-polarization-channel combinations. These should not, however, be interpreted strictly as conventional co- and cross-polarized target returns, since the broadcast illumination may contain mixed polarization and the polarization state of the target-scattered field is not independently characterized. The four combinations are therefore used here to examine polarization-dependent variations in target detection and CAF performance.}  The results presented here demonstrate that all four combinations are capable of detecting and localizing the target.

Fig.~\ref{fig_caf} shows zoomed CAF maps around the expected bistatic excess range and range-rate of the ISS for four polarization combinations: XX, YY, XY, and YX. In each case, a localized peak is observed near the TLE-predicted position (red circle), indicating successful detection of the target in bistatic range–range-rate space. The detected peak locations (black crosses) are in close proximity to the predicted state across all polarization combinations, demonstrating that the system can exploit dual-polarization measurements to support passive detection. Although the peak strength and background structure vary among the polarization pairs, consistent target responses are obtained in all cases, confirming the robustness of CAF-based processing under mixed-polarization broadcast illumination. It should be noted that the CAF maps in Fig.~\ref{fig_caf} correspond to a single snapshot acquired at 24 November 2025, 23:43:19~UTC, consistent with the DoA example (Fig.~\ref{doa_uv_map}) presented earlier. These plots are intended to illustrate instantaneous detection performance and polarization-dependent responses at a representative time during the pass, rather than statistical behavior over the full observation interval. 

{Table~\ref{tab:caf_pol_snapshot} summarizes the bistatic range deviation, bistatic range-rate deviation, and SNR corresponding to the CAF results shown in Fig.~\ref{fig_caf}. These deviations are calculated with respect to the TLE-derived reference trajectory at the snapshot epoch, 24~November~2025 at 23:43:19~UTC.} {The YX combination yields the highest SNR of $24.23~\mathrm{dB}$, whereas the XY combination produces the lowest at $16.03~\mathrm{dB}$, demonstrating that the selected reference--surveillance X/Y channel combination influences the received CAF response. The XX and YY same-polarization-channel combinations achieve intermediate SNR values of $19.70~\mathrm{dB}$ and $20.79~\mathrm{dB}$, respectively.} Moreover, the bistatic range rate deviation, {$\Delta V$}, is consistent across the XX, YY, and XY polarization pairs at $6.95~\mathrm{m/s}$, whilst the YX channel yields a slightly higher value of $9.73~\mathrm{m/s}$. In contrast, the bistatic range deviation, $\Delta R$, varies across the polarization combinations. The XY combination exhibits the smallest deviation from the TLE-derived reference at $0.616~\mathrm{km}$, whereas the YY and YX combinations exhibit the largest deviations, both approximately $0.851~\mathrm{km}$. These differences indicate that polarization selection can influence instantaneous CAF localization performance under broadcast illumination, owing to variations in target aspect geometry, received signal strength, and residual interference.

\begin{table}[t]
\centering
\caption{{Bistatic range and range-rate deviations relative to the TLE-derived reference, together with the SNRs of the four polarization channels, at 23:43:19~UTC on 24~November~2025.}}
\label{tab:caf_pol_snapshot}

\renewcommand{\arraystretch}{1.15}
\begin{tabular*}{\columnwidth}{@{\extracolsep{\fill}}cccc@{}}
\hline
\textbf{Pol.} &
\textbf{{$\Delta R$ (km)}} &
\textbf{{$\Delta V$ (m/s)}} &
\textbf{SNR (dB)} \\
\hline
XX & 0.792 & 6.95 & 19.70  \\
YY & 0.851 & 6.95 & 20.79 \\
XY & 0.616 & 6.95 & 16.03 \\
YX & 0.851 & 9.73 & 24.23 \\
\hline
\end{tabular*}
\end{table}

{Multipath propagation may also contribute to these variations. A recent LOFAR-based study showed that multipath effects depend on the propagation geometry and surrounding reflecting environment, potentially affecting bistatic parameter retrieval \cite{konrad_multipath2026}. Although the geometry is identical for all polarization combinations in the present experiment, single- and multiple-reflection paths may introduce polarization-dependent attenuation, phase shifts, and changes in polarization state. Consequently, the observed differences in SNR and bistatic parameter deviations may result from the combined influence of target scattering, transmitter and receiver antenna characteristics, multipath propagation, residual interference, and calibration uncertainties \cite{jkedrzejewski2025selected}. However, these individual contributions cannot be isolated conclusively from the available single-pass dataset.}

To characterize performance beyond the instantaneous snapshot shown in Fig.~\ref{fig_caf}, CAF detections were extracted over the full 30-s observation interval for all polarization combinations. {The temporal evolution of the bistatic range and range-rate deviations is shown in Fig.~\ref{fig:error_cafs} for all polarization combination channels, with the detection counts indicated in the legend.} Polarization pair YX  achieves the most complete coverage with all 30 detections, followed by YY with 29 out of 30, whilst the XY combination yields the lowest detection count at 25 out of 30 frames. The corresponding detection rates across all polarization channels are summarized in Table~\ref{tab:pol_metrics_core}. All polarization combinations exhibit a similar temporal trend in both {bistatic range and range rate deviations}, though the magnitude of the {deviation} differs between polarization pairs.

\begin{table*}[!t]
\centering
\caption{Core performance metrics for different polarization channels pairs over a 30-second observation interval.}
\label{tab:pol_metrics_core}
\renewcommand{\arraystretch}{1.2}
\setlength{\tabcolsep}{3pt} 
\begin{tabularx}{\textwidth}{l 
>{\centering\arraybackslash}X 
>{\centering\arraybackslash}X 
>{\centering\arraybackslash}X 
>{\centering\arraybackslash}X 
>{\centering\arraybackslash}X 
>{\centering\arraybackslash}X 
>{\centering\arraybackslash}X}
\hline
\textbf{Pol} & \textbf{Det. (\%)} & \textbf{RMSE$_{R}$} (km) & \textbf{STD$_{R}$} (km) & \textbf{RMSE$_{V}$} (m/s) & \textbf{STD$_{V}$} (m/s) & \textbf{Mean SNR} (dB) \\
\hline
XX & 93.33 & 0.953 & 0.416 & 10.88 & 7.35  & 18.44 \\
YY & 96.67 & 1.021 & 0.283 & 10.67 & 7.29  & 20.56 \\
XY & 83.33 & 1.084 & 0.436 & 10.67 & 7.00  & 18.57 \\
YX & 100.0 & 0.988 & 0.319 & 10.85 & 7.18  & 21.14 \\
\hline
\end{tabularx}
\end{table*}

{The bistatic range deviations shown in Fig.~\ref{fig:caf_range} remain predominantly negative across all four polarization combinations, although their magnitude and detection continuity vary with polarization. The consistent negative offset is therefore unlikely to be primarily polarization-specific. However, it cannot be attributed solely to the TLE-derived reference trajectory, as common calibration, propagation, and system-model effects, together with uncertainty in the TLE-derived reference, may also contribute to the observed offset.} {A brief reduction in the bistatic range deviation towards zero is observed across all channels near $t \approx 12-13~\mathrm{s}$, coinciding with the closest point of approach of the ISS.} {After this point, the trend returns to its earlier behavior, with the bistatic range deviation becoming increasingly negative as the pass progresses. This trend may reflect increasing differences between the measurements and the TLE-derived reference, together with other common propagation or system-model effects.}

\begin{figure}[!t]
    \centering
    \begin{subfigure}[b]{\linewidth}
        \includegraphics[width=\linewidth]{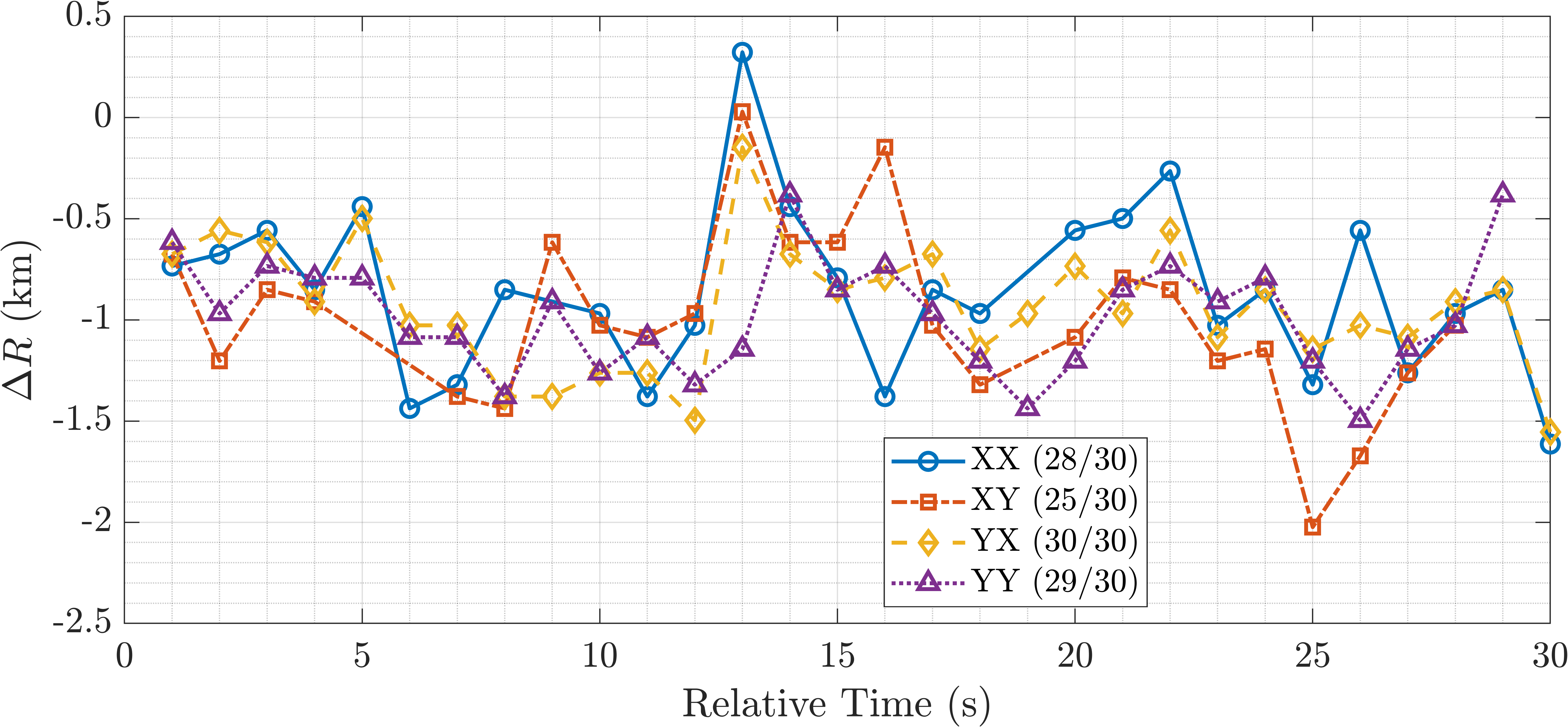}
        \caption{}
        \label{fig:caf_range}
    \end{subfigure}
   \hfill
    \begin{subfigure}[b]{\linewidth}
        \includegraphics[width=\linewidth]{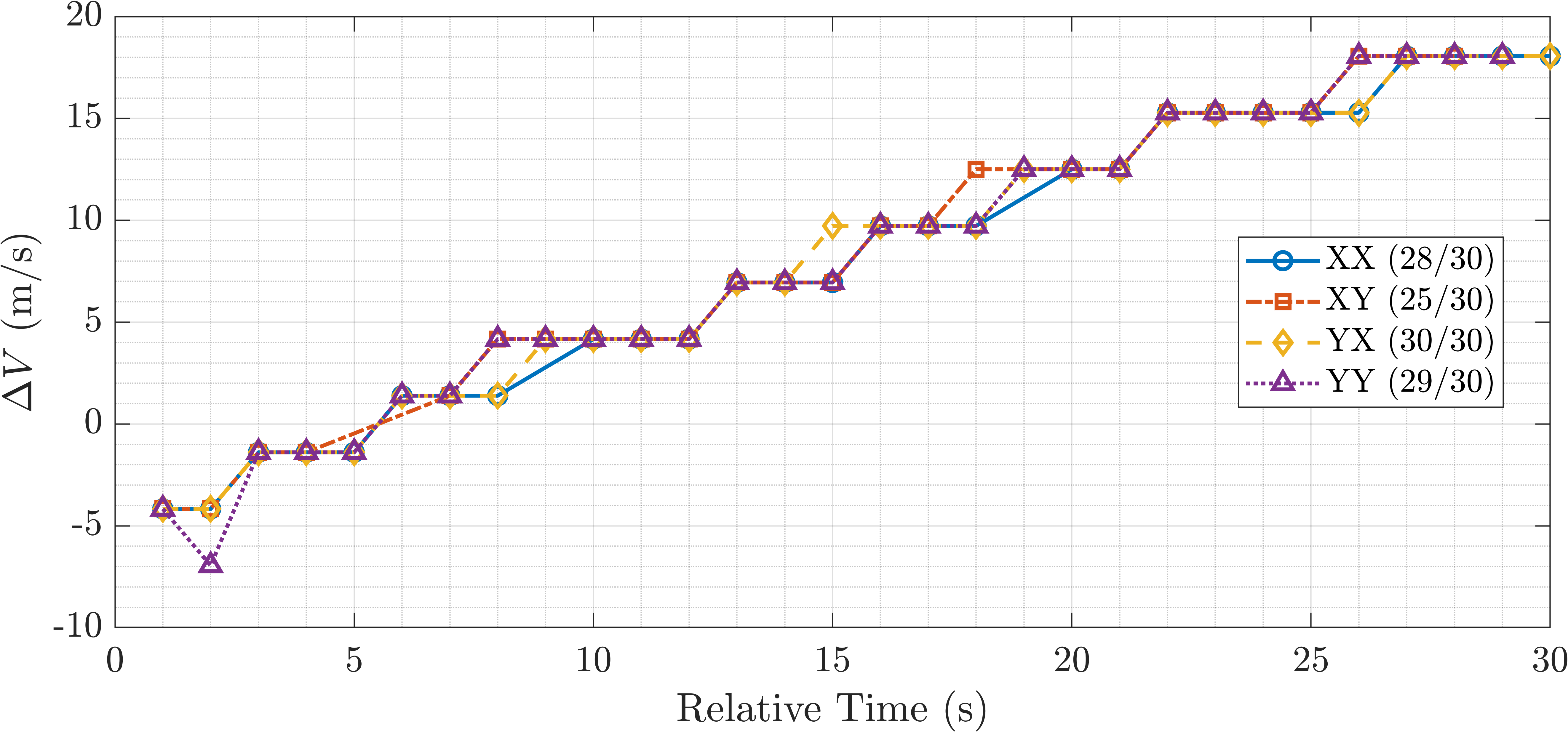}
        \caption{}
        \label{fig:caf_vel}
    \end{subfigure}
    \caption{Bistatic parameter {deviations} relative to the TLE-predicted reference over a 30-second observation pass for each transmit-receive polarisation pair: (a) bistatic excess range and (b) bistatic range-rate.}
    \label{fig:error_cafs}
\end{figure}

Fig.~\ref{fig:caf_vel} shows the corresponding bistatic range-rate {deviations} for the observation pass. In contrast to the range estimates, the range-rate measurements are highly consistent across polarization pairs, reflecting the dominant influence of CPI duration and target motion on Doppler estimation. The increasing range-rate deviation over time is likely due to residual mismatch between the TLE-based motion compensation and the true target dynamics. As can be seen in Fig.~\ref{fig:caf_vel}, the range rate {deviation} exhibits periodic step-like jumps, with each jump corresponding to a single Doppler/range rate resolution cell. For a CPI duration of $1~\mathrm{s}$ at \(107.9~\mathrm{MHz}\), the range rate resolution is \(\Delta~v= \lambda/T_{CPI} \approx 2.78~\mathrm{m/s}\), which is consistent with the step size observed in Fig.~\ref{fig:caf_vel}. {The step-like behaviour is mainly associated with the discrete Doppler grid, for which the estimated range rate changes by one resolution cell as the residual target Doppler crosses a bin boundary. Such residual offsets may arise when the TLE-based compensation does not fully account for higher-order delay and Doppler migration over the CPI. Previous MWA studies have shown that Doppler-rate compensation and the Keystone transform can mitigate Doppler and range walk, respectively, thereby improving coherent-integration sensitivity \cite{hennessy2023surveillance}. The implementation and evaluation of these higher-order motion-compensation techniques are left for future work.}

To quantify the overall performance, the metrics computed across all detected frames are summarized in Table~\ref{tab:pol_metrics_core}. The YX channel combination achieves the highest detection rate of 100\%, followed by YY at 96.67\%, XX at 93.33\%, and XY at 83.33\%, demonstrating a dependence of detection performance on the selected reference--surveillance X/Y channel combination. {In terms of bistatic range deviation, the RMSE relative to the TLE-derived reference ranges from approximately $0.95$ to $1.08~\mathrm{km}$ across the tested configurations, with standard deviations between $0.28$ and $0.43~\mathrm{km}$.} The YY configuration achieves the lowest variability ($\mathrm{STD}_R = $0.28$~\mathrm{km}$), while XY shows the largest RMSE ($1.08~\mathrm{km}$). It is important to note that the  {achievable bistatic range resolution} is fundamentally limited by the bandwidth of the transmitted signal. For the FM broadcast signal used in this experiment, the effective bandwidth varies with the instantaneous programme content and can approach the nominal $200~\mathrm{kHz}$ channel bandwidth. Using $\Delta R_{\mathrm{b}}=c/B$, an effective bandwidth of 200~kHz corresponds to a best-case bistatic excess-path resolution of approximately $1.5~\mathrm{km}$ \cite{malanowski2019signal}. In contrast, the bistatic range-rate RMSE is consistent across all four channels, lying between $10.67~\mathrm{m/s}$ and $10.88~\mathrm{m/s}$, with corresponding STDs between $7.00~\mathrm{m/s}$ and $7.35~\mathrm{m/s}$. {The uniformity of the Doppler/range-rate deviation across polarization channels further supports the interpretation that range-rate estimation is influenced primarily by the CPI duration, motion compensation, calibration offsets, and other common system-model effects rather than the polarization configuration.}

{Another metric of interest is the CAF-output target SNR across all detected frames in the 30-s observation interval. Fig.~\ref{fig:snr_box} summarizes its distribution for each reference--surveillance polarization combination using box plots, while the corresponding mean values over the complete pass segment are reported in Table~\ref{tab:pol_metrics_core}. The YX combination provides the highest mean SNR of $21.14~\mathrm{dB}$, followed by YY at $20.56~\mathrm{dB}$, XX at $18.44~\mathrm{dB}$, and XY at $18.57~\mathrm{dB}$. The boxplots further show that YX and YY exhibit higher median SNRs, whereas XX and XY provide lower but broadly comparable distributions. As shown in Fig.~\ref{fig:snr_box}, the measured CAF-output SNRs are lower than the nominal $26-29~\mathrm{dB}$ range obtained from the link-budget analysis in Section~\ref{subsec:link_budget}. This difference reflects the sensitivity of the first-order prediction to assumptions regarding the effective bistatic RCS, transmitter illumination, receive-array gain, system noise temperature, and aggregate losses. The predicted SNR therefore serves as a nominal reference for comparison with the measurements.}

The observed differences in SNR across the polarization combinations are also reflected in the detection continuity and the stability of the bistatic range and range-rate estimates. The Y-reference configurations, YX and YY, achieved detection probabilities of 100\% and 96.67\%, respectively, whereas the X-reference configurations exhibited more missed detections and larger deviations. {Since the X- and Y-polarized reference beams exhibit comparable SNR levels and low temporal variability, as discussed in Section~\ref{subsec:link_budget}, these differences are unlikely to arise solely from the quality or stability of the reference signal. Instead, they may result from polarization-dependent echo strength and residual interference, together with the scattering characteristics of the ISS. Its large and complex structure, including the truss, modules, and solar arrays, forms multiple scattering centers. As the bistatic aspect changes during the pass, the coherent combination of returns from these structures may produce polarization-dependent RCS fluctuations and intermittent constructive or destructive scattering. Together with the propagation effects discussed earlier, these variations can lead to short-term changes in echo strength and reduced detection continuity.}

\begin{figure}[t]
    \centering
    \includegraphics[width=0.95\linewidth]{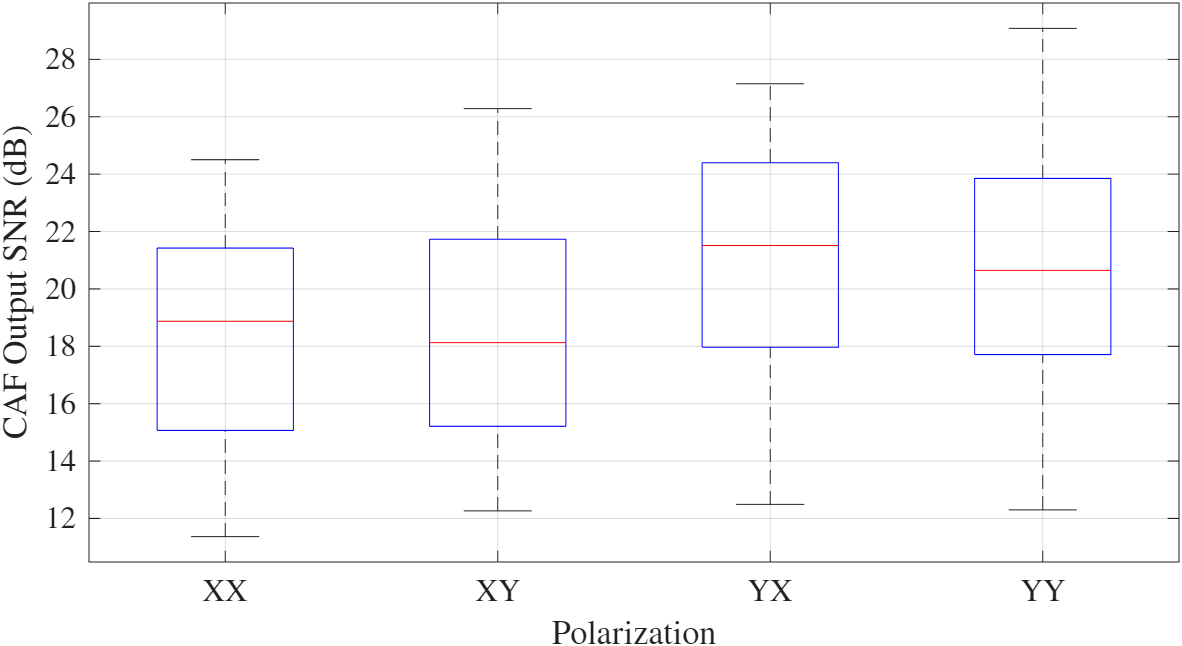}
    \caption{SNR distribution for each transmit-receive polarization combination over the 30-second observation pass.}
    \label{fig:snr_box}
\end{figure}

{Another factor influencing the interpretation of the reported performance metrics is the uncertainty of the propagated TLE used as the reference trajectory. For typical low-Earth-orbit satellites with reasonably accurate TLEs, SGP4 propagation over 1--2 days generally results in radial and cross-track uncertainties on the order of a few hundred metres, {whereas the along-track uncertainty is typically larger and may reach several kilometres. The along-track component is generally the dominant source of orbit prediction uncertainty and increases rapidly with propagation time \cite{aida2013accuracy}.} Consequently, the reported angular, bistatic range, and bistatic range-rate deviations inherently reflect the uncertainty of the propagated reference trajectory and should therefore be interpreted as agreement with the propagated TLE rather than absolute errors with respect to the true target trajectory.}

{Beyond the uncertainty associated with the propagated TLE reference, the observation geometry also plays an important role in the obtained results.} For the Adelaide FM transmitter, located at a baseline distance of approximately $220~\mathrm{km}$ from the receiver, the ISS attains relatively high elevation angles with respect to the transmitter during portions of the pass. Under these conditions, the target may lie within lower-gain regions of the broadcast antenna pattern, reducing the effective illumination strength. Moreover, this FM broadcast transmitter in Adelaide radiates in mixed polarization to serve both horizontally polarized domestic antennas and vertically polarized vehicle antennas, distributing transmitted power across both polarization planes. {Consequently, signal energy can be present in both X and Y receive channels of the InteRFeX array, enabling the four reference--surveillance X/Y channel combinations considered in this work.} {Additionally, the approximately $220~\mathrm{km}$ transmitter--receiver separation increases the susceptibility of the reference path to terrain blockage and multipath, potentially attenuating the reference signal and reducing the CAF-output SNR. These effects could be mitigated through a dedicated directional antenna for reference-channel acquisition.}

{A limitation of the present experimental validation is that it is based on a single ISS pass, with the analysis restricted to a $30~\mathrm{s}$ high-elevation observation interval. Consequently, the reported performance is representative of this specific observation geometry and does not establish system performance over longer tracking intervals or across different passes and elevation angles. Further experiments using longer observation arcs and multiple passes are therefore required to assess the robustness and sustained tracking performance of the system under a broader range of observation conditions.} Moreover, it should be emphasized that these results are preliminary. Although basic calibration procedures have been applied, refinement is ongoing, and residual biases may still be present in the reported measurements. Nevertheless, the experimental results demonstrate that InteRFeX is capable of providing consistent bistatic and angular measurements of LEO satellites using FM broadcast illumination, thereby establishing a foundation for more extensive experimental validation. The polarization-dependent analysis further indicates that polarization selection can influence detection continuity under practical broadcast conditions, motivating further investigation of dual-polarization reception.

\section{Conclusion}\label{conclusions}
This paper described and experimentally validated a passive radar signal processing pipeline for InteRFeX, exploiting commercial FM transmitters as illuminators of opportunity for LEO space object detection. The work demonstrated a dedicated passive radar architecture for Space Domain Awareness, showing how radio-astronomy array capabilities can be adapted for the detection and monitoring of space objects. System performance was evaluated through MUSIC-based direction-of-arrival estimation and cross-ambiguity function processing, yielding angular localization and bistatic range and range-rate measurements of the target. {The ISS trajectory was validated over the observation interval by comparing the estimated measurements with the corresponding TLE-based predictions.} {The results showed a mean angular separation of approximately $0.45^\circ$, with bistatic range and range-rate RMSEs of $0.953$--$1.084~\mathrm{km}$ and $10.67$--$10.88~\mathrm{m/s}$, respectively, indicating consistent deviations from the TLE-derived reference across the polarization combinations. These deviations may reflect the combined influence of system-calibration uncertainties, array-pointing misalignment, motion-compensation limitations, propagation effects, and other model uncertainties. Further investigation using multiple satellite passes and observation geometries is required to better isolate and quantify the contributions of these effects.}

{In addition, decomposition of the angular deviation into track-frame components indicated no appreciable systematic timing offset, while the mean cross-track bias of $0.38^\circ$ revealed a systematic angular offset relative to the TLE-derived reference. The phase-gradient analysis identified residual array phase-calibration errors as a plausible contributor, although array-pointing misalignment and other model errors may also contribute.} Despite that, these results {show} that both angular and bistatic measurements can be extracted from passive FM-based observations, even in the presence of interference and competing astronomical sources. 

{The availability of the orthogonal X/Y receive channels further enabled a polarization-dependent assessment of detection performance.} The results suggest that polarization pairing can influence detection continuity and {the reliability of bistatic range and range-rate estimation} under the considered broadcast conditions. Across the four reference--surveillance polarization combinations, the YX channel achieved the highest detection rate of 100\% and the highest mean SNR of $21.14~\mathrm{dB}$, whereas the XY combination yielded the lowest detection rate of 83.33\% and a mean SNR of $18.57~\mathrm{dB}$. {These measured SNR levels are below the approximately $29$--$26~\mathrm{dB}$ range estimated from the link-budget analysis, indicating that the first-order model does not capture all effects present in the experiment. This motivates a more detailed characterization of the propagation, scattering, and system parameters in future measurements.} Overall, the demonstrated measurements provide a foundation for developing cost-effective passive radar systems for continuous monitoring and tracking of LEO objects, with the presented signal processing pipeline serving as a basis for enabling a dedicated passive radar mode within the InteRFeX system.

\section{Future Work}\label{future_work}
{Future work will focus on refining the signal-processing framework and calibration procedures, together with validation over longer observation intervals, additional target passes, and a wider range of bistatic geometries. In addition, the present work forms the reference and surveillance channels using the same antenna array to reduce hardware complexity and avoid inter-channel synchronization requirements. Future studies will investigate the use of a dedicated reference channel where improved direct-path signal quality is beneficial. Moreover, simultaneous observations using multiple spatially separated FM transmitters will be explored to increase geometric diversity.} The extracted bistatic range, range-rate, azimuth, and elevation measurements will be used within an orbit determination and tracking framework to support continuous tracking and improved state estimation of LEO objects. Moreover, future work will investigate the coherent fusion of dual-polarization channels to exploit the full polarimetric information available from the array, with the aim of improving detection reliability for space surveillance applications.

%

\section*{Acknowledgment}
The authors thank Nova  Systems for supporting this work and facilitating the system at the Nova Systems Space Precinct, Peterborough, South Australia. This paper and related research have been conducted during and with the support of the Italian National inter-university PhD program in Space Science and Technology. 

\section*{Appendix}
The complete set of processed 1-s Bistatic range--range rate frames for all polarization combinations is provided as supplementary material at the following link: \url{https://drive.google.com/drive/folders/1egg-vXCnIZp8Mj_9J3KY05q_wiIP9DIn?usp=drive_link}

\bibliographystyle{IEEEtran}
\bibliography{reference}   

\begin{IEEEbiography}[{\includegraphics[width=1in,height=1.25in]{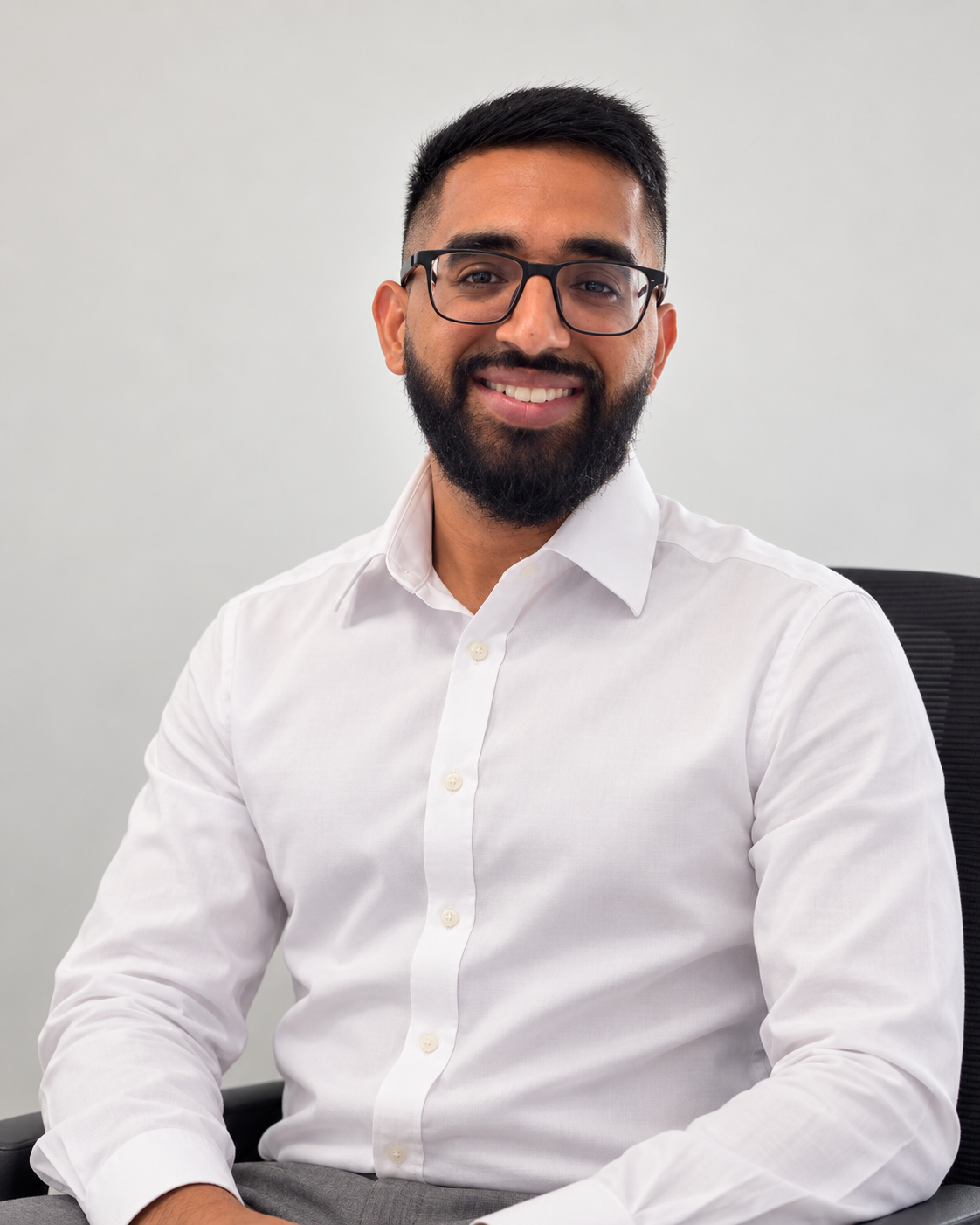}}]{Bhaskar Ahuja}
received the B.Sc. (Hons.) and M.Sc. degrees in physics from the University of Delhi, India, in 2018 and 2020, respectively, and the Ph.D. degree in Space Science and Technology from the University of Trento, Italy, in 2026. His doctoral research, carried out in collaboration with the RaSS Laboratory, CNIT, focused on radar-based Space Domain Awareness, particularly the detection, tracking, and characterization of space objects. During his Ph.D., he was a visiting researcher at Curtin University, Perth, Australia, working on radar signal processing and digital beamforming. Previously, he was a Research Fellow with the Indian Institute of Remote Sensing, ISRO, India. His research interests include radar systems, passive radar, beamforming, space surveillance and tracking, orbit determination, and signal processing for space applications. 
\end{IEEEbiography}
\begin{IEEEbiography}[{\includegraphics[width=1in,height=1.25in,clip,keepaspectratio]{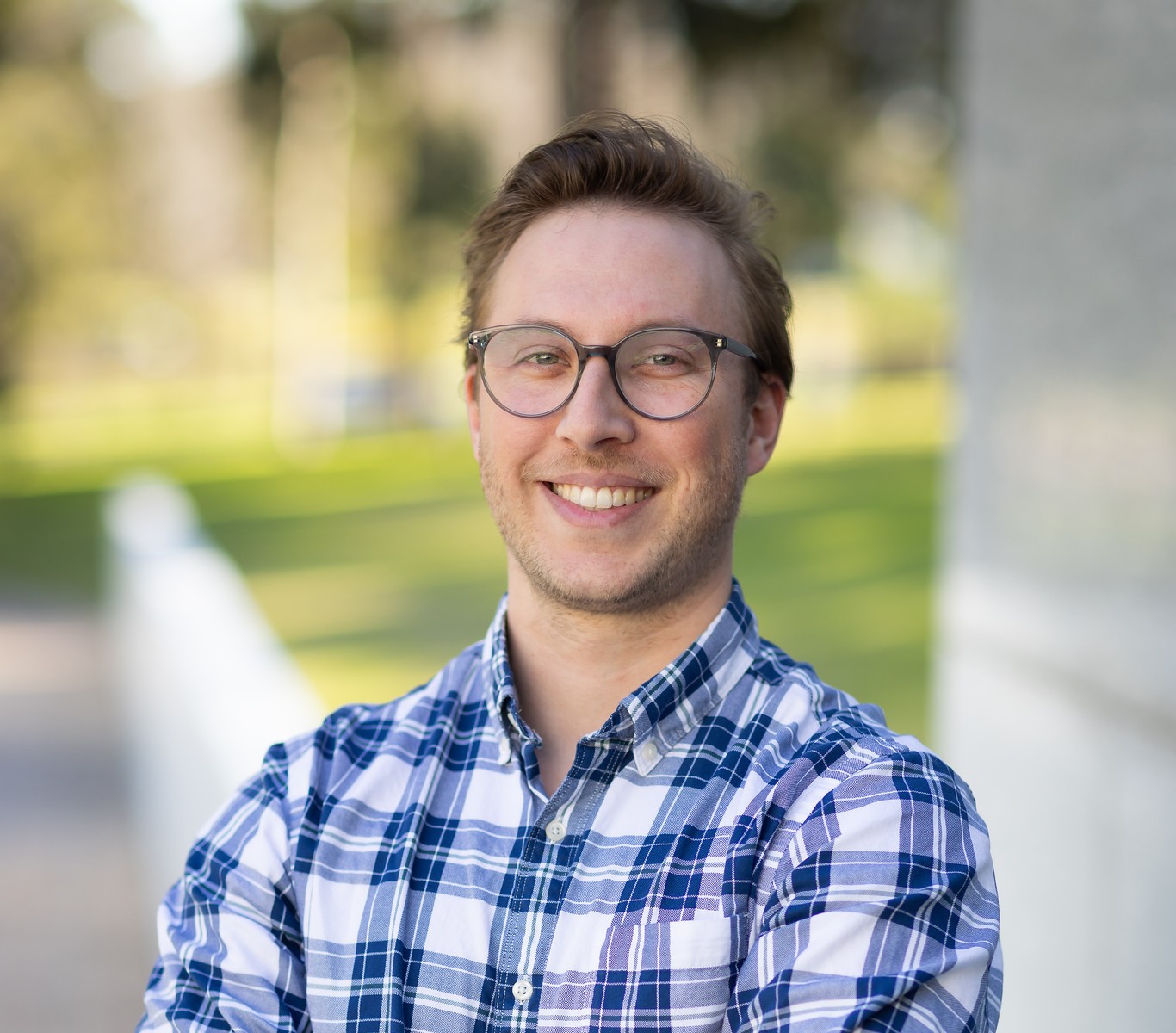}}]{Thomas Crowther}
received a Bachelor of Mechatronic Engineering from Curtin University, Australia, in 2014, and is currently completing a PhD in Engineering at the Curtin Institute of Radio Astronomy. His doctoral research, conducted in collaboration with Nova Systems, focuses on the development of InteRFeX, a purpose-built, low frequency passive radar array dedicated to Space Domain Awareness. His research areas include the use of radio astronomy techniques to Space Domain Awareness, instrumentation and engineering, and algorithmic automation for data processing.
\end{IEEEbiography}
\begin{IEEEbiography}[{\includegraphics[width=1in,height=1.25in,clip,keepaspectratio]{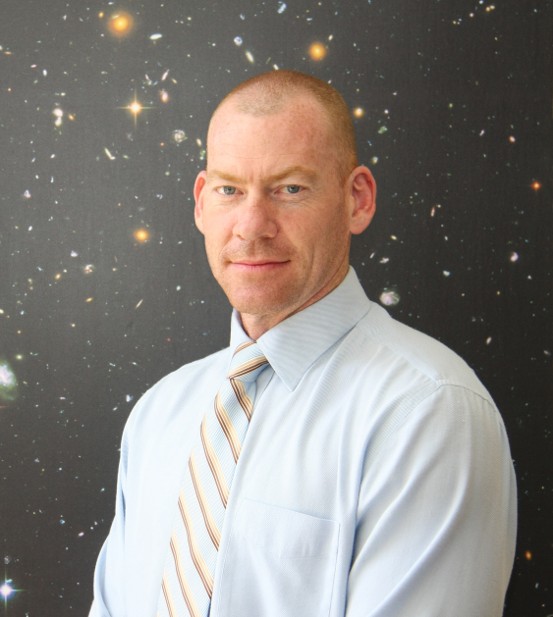}}]{Steven Tingay}
received an Honours degree in Physics from The University of Melbourne in 1991, and his PhD in Astronomy and Astrophysics from the Australian National University in 1997.  He is now a John Curtin Distinguished Professor at Curtin University, where he is the Executive Director of the Curtin Institute of Radio Astronomy (which he founded in 2007), Director of the Murchison Widefield Array, and Deputy Executive Director of the International Centre for Radio Astronomy Research.  He is a radio astronomer, with interests across astrophysics, instrumentation and engineering, and data processing with high performance computing.  He has an interest in the translation of astronomy technologies and techniques into other application areas, such as space and defence.  Prior to his position at Curtin University, he held positions at Swinburne University of Technology (2003 – 2007), CSIRO (1999 – 2003), and the Jet Propulsion Laboratory (1996 – 1999).

\end{IEEEbiography}
\begin{IEEEbiography}[{\includegraphics[width=1in,height=1.25in,clip,keepaspectratio]{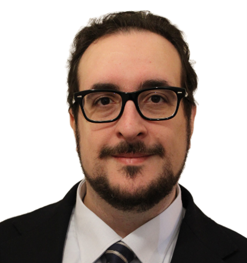}}]{Luca Gentile}
Luca Gentile received his B.Sc. degree in Telematics Engineering from Kore University of Enna, Italy, in 2012, and his M.Sc. degree in Telecommunications Engineering from the University of Pisa, Italy, in 2016. He received his Ph.D. in Information Engineering from the University of Pisa in 2020 and, during his Ph.D., he spent six months in 2018 conducting research at The University of Texas at Austin, USA.
He is currently a Researcher at the RaSS National Laboratory at Consorzio Nazionale Interuniversitario per le Telecomunicazioni (CNIT) in Pisa, Italy. His research focuses on radar systems, radar target tracking, space surveillance, and space situational awareness.

\end{IEEEbiography}

\begin{IEEEbiography}[{\includegraphics[width=1in,height=1.25in,clip,keepaspectratio]{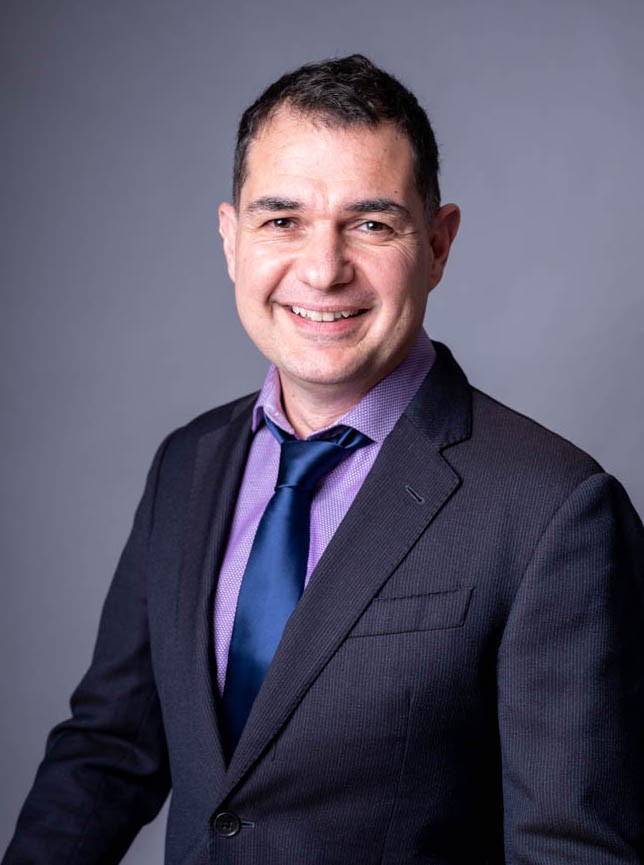}}]{Marco Martorella}
received his Laurea degree (Bachelor+Masters) in Telecommunication Engineering in 1999 (cum laude) and his PhD in Remote Sensing in 2003, both at the University of Pisa. He is now Professor at the School of Engineering of the University of Birmingham, where is Head of the Microwave Integrated Systems Laboratory (MISL), Director of the Communications and Sensing Cluster and Director of the Space Communications and Sensing Laboratory. He is also Vice-Director at the Radar and Surveillance Systems (RaSS) National Laboratory in Pisa (Italy). He was Chair of the NATO Sensors and Electronics Technology (SET) Panel in the years 2023-2025 and Italian Academic Delegate within the European Defence Agency in the years 2018-2022. He is author of about 300 international journal and conference papers, 4 books and more than 20 book chapters. He has been recipient of the IEEE 2013 Fred Nathanson Memorial Radar Award, the 2016 Outstanding Information Research Foundation Book publication and three NATO STO Excellence Award (2022, 2024, 2026). He is a co-founder of ECHOES, a radar systems-related spin-off company. His research interests are mainly in the field of radar, with specific focus on radar imaging, multidimensional radar, passive radar and space situational awareness. He is a Fellow of the IEEE.
\end{IEEEbiography}

\end{document}